\documentclass{aa}

\usepackage{graphicx}
\usepackage{txfonts}
\usepackage{ wasysym }
\usepackage{float}
\usepackage{verbatim}
\usepackage[title]{appendix}
\graphicspath{{imagenes_resultados/}}
\usepackage{xcolor}
\usepackage[normalem]{ulem}
\usepackage{subcaption}

\begin{document}

   \title{The relationship between nuclear rings and the overall galaxy morphology}

   \author{P. Jim\'enez-S\'anchez
          \inst{1}\fnmsep\inst{2}
          \and S. Comer\'on \inst{2}\fnmsep\inst{3}
          \and A. Prieto \inst{3}\fnmsep\inst{2}\fnmsep\inst{4}
          \and J. H. Knapen \inst{3}\fnmsep\inst{2}
          }
   
   \institute{Institut de Fisica d’Altes Energies (IFAE), The Barcelona Institute of Science and Technology, Campus UAB, 08193 Bellaterra (Barcelona), Spain
         \and   
            Departamento de Astrofísica, Universidad de La Laguna, 38200 La Laguna, Tenerife, Spain
         \and
             Instituto de Astrofísica de Canarias, 38205 La Laguna, Tenerife, Spain
        \and
             Universitats-Sternwarte M\"unchen, D-81679 M\"unchen, Germany
             }

   \date{Received May 9, 2025; accepted June 27, 2025}

  \abstract
   {Nuclear rings are composed of gaseous material and/or stars and have a typical size ranging from 100s pc to $\sim$ 1 kpc. Their study is crucial in unravelling the processes involved in the evolution of the central regions of galaxies. Nuclear rings are long-lived structures, allowing the molecular gas inside to become sufficiently dense to initiate star formation. This makes them contributors to young circumnuclear populations and thus a crucial element in the study of secular evolution. However, the morphology of nuclear rings, and their potential correlations with that of the galactic hosts, remains an open subject.}
   {We examine 52 star-forming nuclear rings from the Atlas of Images of NUclear Rings (AINUR) and correlate the overall galaxy morphology, in particular non-axisymmetric features, with the morphology of the nuclear ring.}
   {We divide the sample into different classes according to two visual classifications, one based on the overall morphology of the galaxy, and another one based on the morphology of the nuclear ring. We define three different classes of nuclear rings: two-armed rings dominated by two dust lanes, twoarms+ rings crowded with secondary dust lanes in addition to the two main ones, and many-armed rings with multiple armlets of similar prevalence. We employ unsharp-masked \textit{Hubble Space Telescope} images to study the structure of nuclear rings.}
   {We find that two-armed rings are more common in early-type grand design galaxies with strong bars. Twoarms+ rings are related to later-type and more weakly barred galaxies, both grand design and multi-armed. Lastly, many-armed rings are typically associated with later-type flocculent and multi-armed galaxies with the weakest bars. In addition, we examine the regions inside the nuclear rings and observe nuclear spirals in 28 galaxies ($\sim 90\%$ of those galaxies for which the interior of the nuclear ring is resolved).}
   {We conclude that the global morphology of the host galaxy and, more precisely, the presence and properties of a bar play a fundamental role in determining the morphology of the nuclear ring and the nuclear region. Specifically, the strength of non-axisymmetries strongly influences the morphological characteristics of a galaxy, from its innermost to its outermost regions. We suggest that two-armed rings are associated with a 180º structure forced by a strong bar, many-armed rings are associated with a very weak or absent 180º structure, and twoarms+ rings are found in intermediate cases.}


   \maketitle
\section{Introduction}

  Nuclear or circumnuclear rings are observed in the central regions of certain galaxies, frequently exhibiting enhanced star-formation activity. They are defined as closed, or nearly closed, circular, or almost circular structures composed of gaseous material and/or stars with a typical size  ranging from 100s pc to $\sim$ 1 kpc and a relative size of $\sim0.06~D_{25}$\citep{Laine_2002}, where $D_{25}$ is the standard isophotal diameter defined by a $B$-band surface brightness of $\rm 25~mag~arcsec^{-2}$. 

 \begin{figure*}[h!]
   \centering
   
   \includegraphics[height=5.62 cm]{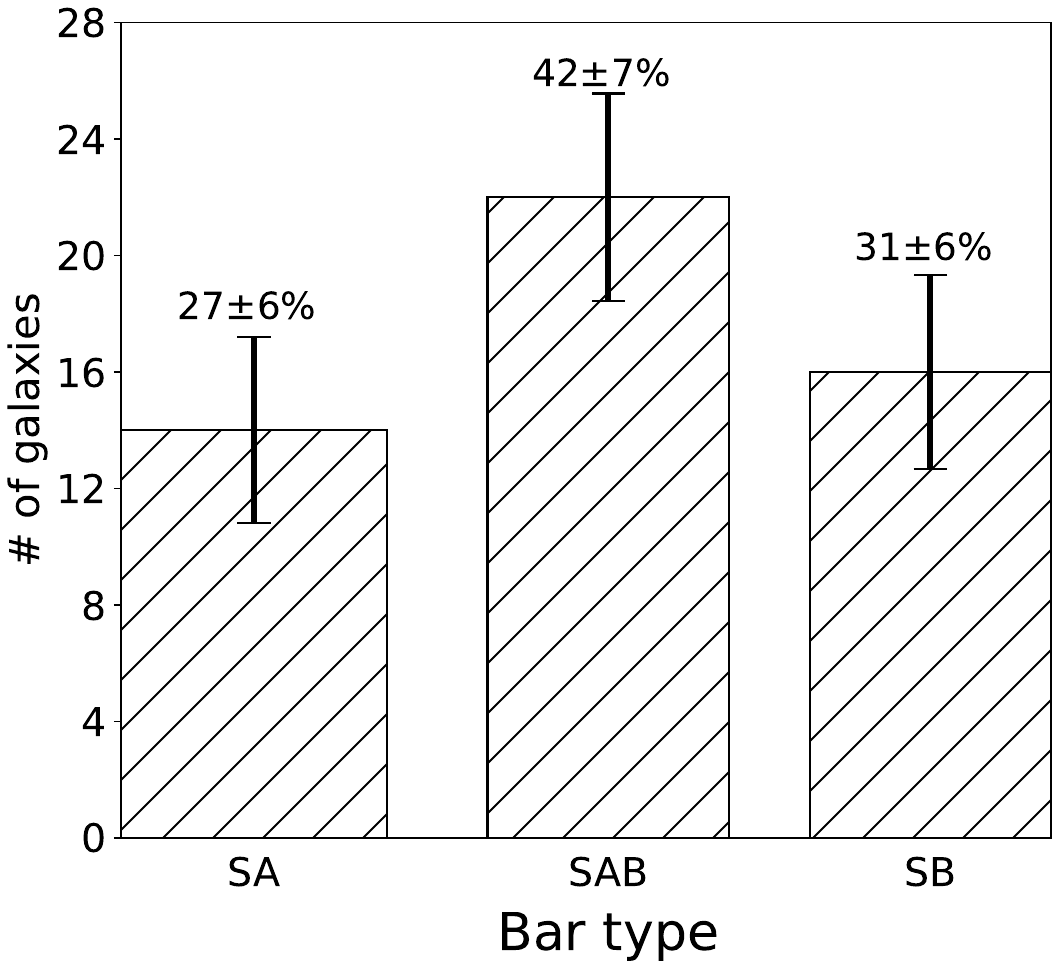}
   \includegraphics[height=5.62 cm]{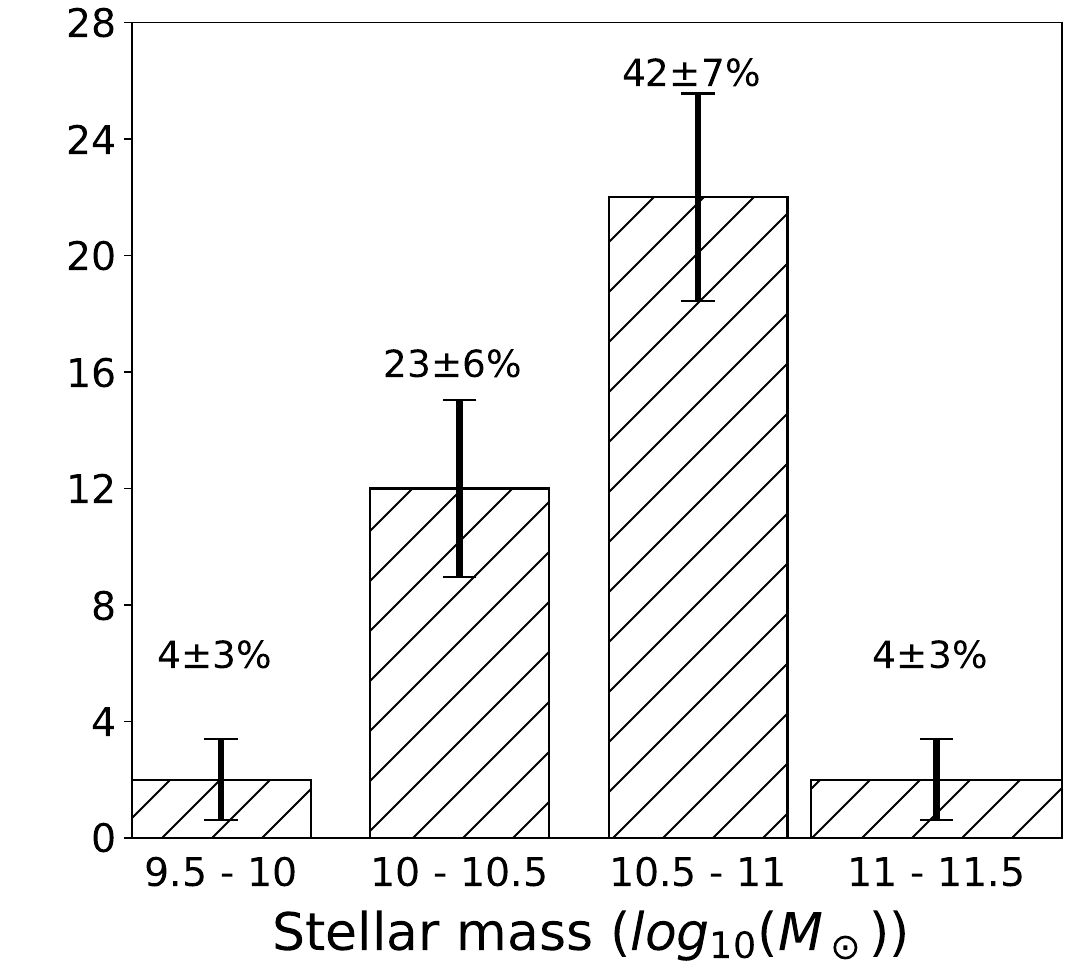}
   \includegraphics[height=5.62 cm]{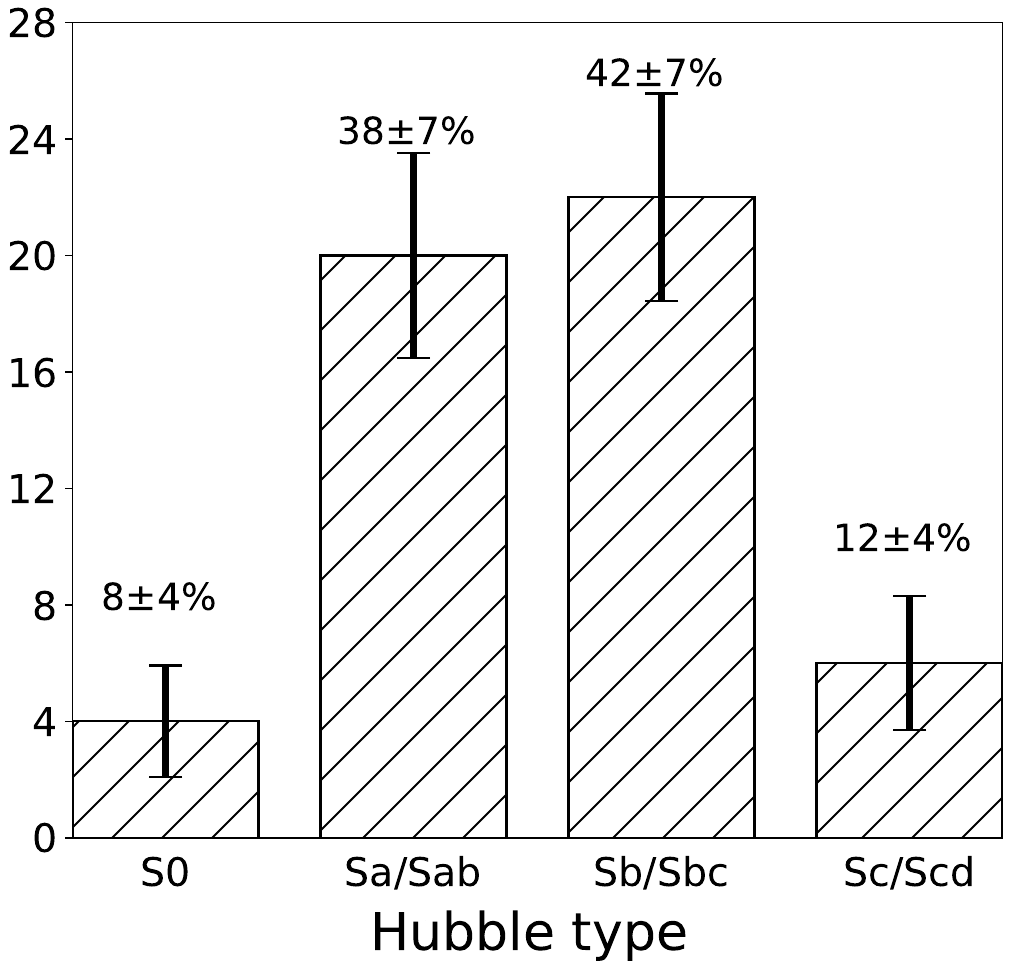}
      
   \caption{Distribution with bar type, stellar mass and Hubble type for our sample of galaxies. The percentage of galaxies belonging to each category, along with  an uncertainty assigned assuming a binomial distribution, is shown for each case. Data from Table \ref{tab:tablafiltros}.  }
              \label{tododist}
    \end{figure*}

  The origin of nuclear rings is thought to be linked to an inflow of gaseous material towards the galactic centre. Nuclear rings thus act as tracers of recent gas inflow to the circumnuclear region \citep[see e.g.][]{KNAPEN1995, PRIETO2005,PRIETO2019,prieto2024}. This streaming motion can be a consequence of a loss of angular momentum \citep[see e.g.][]{COMBESGERIN1985} caused by the presence of non-symmetric morphological features such as bars \citep[see e.g.][]{SHLOSMAN1990,ATHANA1994,COMBES2001}. According to the leading nuclear ring formation hypothesis, after dynamical shocks carry the gas towards the centre following {elongated} orbits, the material crosses an Inner Lindblad Resonance and its trajectory begins to shift from {elongated} to circular {\citep[see][]{ENGLMAIER1997,KNAPEN1995,Knapen_2002}}. As proposed by \cite{COMBES1996} {there is a connection between the location of nuclear rings and the ILRs;  nuclear rings appear in the region} between the outer and inner ILRs in galaxies that exhibit both, and {inside} the single ILR in galaxies where only one of these features is present {\citep[see e.g. the review by][]{SHLOS1999}}. However, numerical results from \cite{KIM2012} suggest that { rings are formed as a result of the centrifugal barrier that the migrating material encounters and later shrink} in size as gas with lower angular momentum is added. \cite{KIM2012} claim that the location of the ring between two ILRs is a coincidence \citep[see also][]{REGAN}. The exact radius (relative to the ILRs) at which a nuclear ring may form is still not a closed subject \citep[see e.g.][]{VENCHANG2009}. It has also been suggested that nuclear rings {correspond to} the edge of a larger nuclear disk. For example, \cite{Bittner_2020} argue that the origin of these nuclear disks, and hence the nuclear rings, is an inside-out formation from a series of small bar-built rings that grow in size as the bar evolves.
  
  Nuclear rings are very stable and long-lived structures \citep{REGAN,KNAPEN2006}, which allows the molecular gas accumulating within them to become sufficiently dense to initiate star formation. This makes them useful tracers of several generations of star formation in the circumnuclear {region} and thus a crucial element in the study of secular evolution {there} \citep[]{COMERON2008b,PRIETO2019}. Furthermore, they play a pivotal role in the evolution of barred galaxies, frequently functioning as a site of active star formation in proximity to the galactic centre \citep[see][]{BUTA1986,GARCIA1991,BARTH1995,MAOZ2001,MAZZUCA2008}, {although circumnuclear rings with active star formation are not always associated with bars} \citep[e.g.][]{ontiveros2009, prieto2024}. Therefore, they contribute to the secular growth of pseudo-bulges \citep[e.g.][]{2006Kormendy}.

This work is conceived as a continuation of the work by \cite{AINUR}, the Atlas of Images of NUclear Rings (AINUR). The AINUR included 113 rings found in 107 galaxies, and a detailed analysis of these rings that aims to explore possible relationships between the size and morphology of the rings and various galactic parameters. 

The AINUR followed the path of previous Atlases, the likes of \cite{SERSIC1965}, \cite{POGGE1989}, \cite{BUTACROC1993}, \citet*{KNAPEN2002}, and  \cite*{KNAPEN2005}, but taking advantage of the high resolution provided by the \textit{Hubble Space Telescope (HST)} to include the intermediate and small-sized nuclear rings. It presented pioneering work on a uniform and comprehensive study of every known nuclear ring, as previous efforts had focussed on individual or small groups of rings. The galaxy sample in the AINUR is mainly based on the survey by \cite{HO1995} {aimed at finding} ‘dwarf’ Seyfert nuclei in nearby galaxies to check for correlations between the ring properties and active galactic nuclei (no correlation was found). The sample was later expanded with a selected group of nuclear rings from the literature. 

The authors of the AINUR conclude that star-forming nuclear rings are found in $20 \pm 2$ per cent of disc galaxies in the range of Hubble morphological types $-3 < T \leq 7$, implying an effective nuclear ring lifetime of $2-3$ Gyr, in general agreement with a wide range of other observational and numerical studies. Furthermore, the authors found that the size of nuclear rings is constrained by the size and ellipticity of the bars, a correlation later confirmed by \cite{ERWIN2024}. In addition, it was observed that star-forming rings are predominantly found in galaxies with morphological types ranging from S0 to Sc, without strong preference for barred host galaxies. 

This paper aspires to expand upon the AINUR by taking a closer look at star-forming nuclear rings, their differences from one another, and the properties of their host galaxies that may cause these differences. To achieve this, the images of the nuclear rings {have} been enhanced through the use of the unsharp-masking technique. This facilitates the differentiation of distinct classes of rings based on their structural characteristics and the subsequent examination of correlations with the known properties of the hosts.

  The study of nuclear rings is of crucial importance in the unveiling of the processes that take place in their formation and in the evolution of the central region of galaxies. The aim of this study is to establish correlations between the overall galactic morphology, in particular non-symmetric features, and the structure of the nuclear rings.

\section{Sample and image selection}
The sample consists of galaxies that exhibit star-forming nuclear rings, derived from the complete original sample of 113 nuclear rings from the AINUR \citep{AINUR}. Of these, 105 exhibited star formation, 84 of which have been observed with the \textit{HST} and were selected for potential analysis of their structure. We {only select galaxies} with \textit{HST} images to study the inner structure of these objects with high angular resolution. The sizes of nuclear rings are usually within the range of hundreds of parsecs, and in consequence their study demands the use of images with an angular resolution reaching tens of parsecs, only available for space telescopes and telescopes with adaptive optics, {even for nearby galaxies}.

Of the 84 star-forming nuclear rings observed by \textit{HST}, 32 were excluded due to lack of clarity within the image, insufficient coverage of the desired object, or the unavailability of images in the optical range. The final sample is composed of 52 galaxies.

   \begin{figure*}[h!]
   \centering
 
  \includegraphics[width = \textwidth]{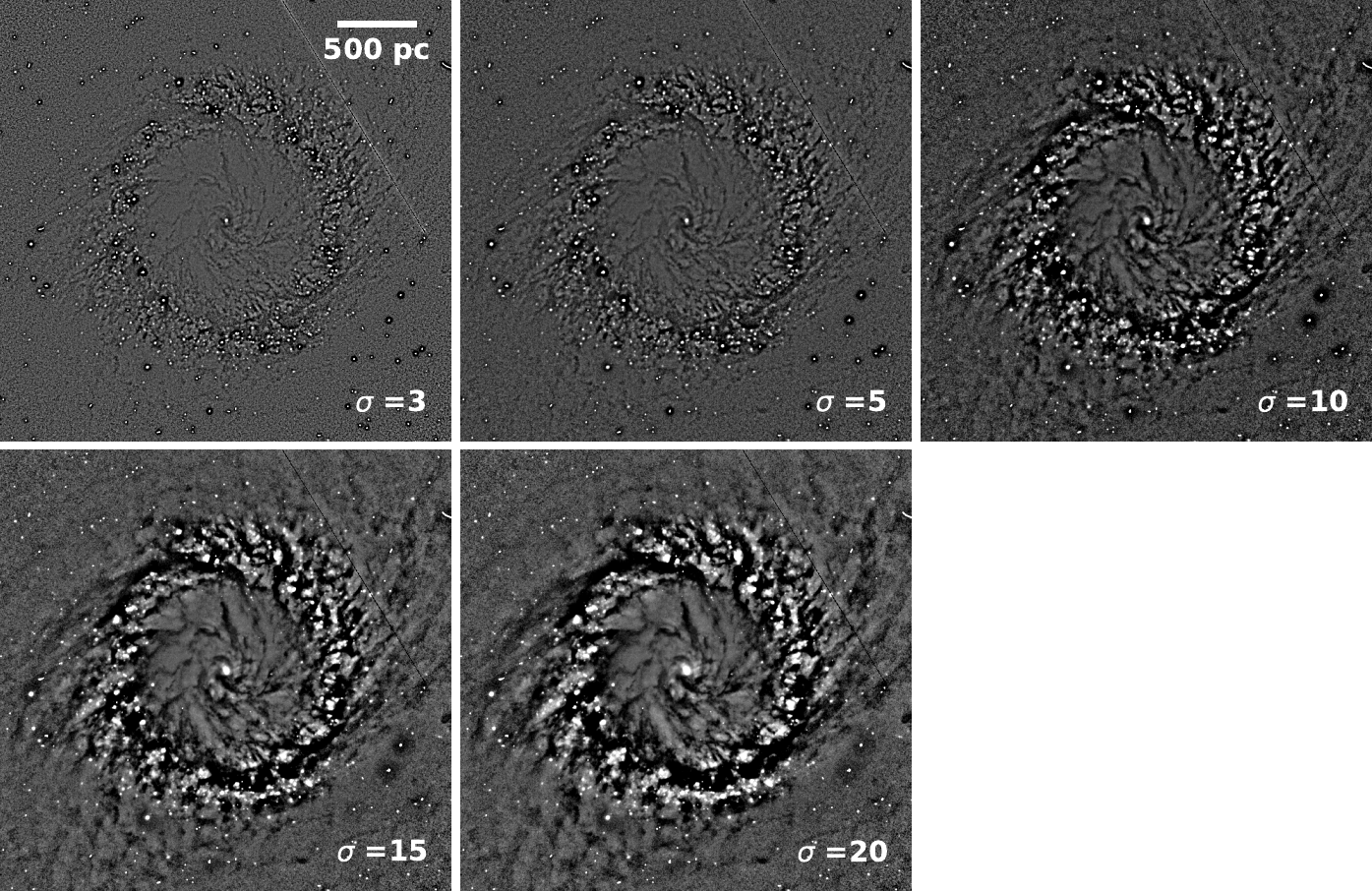}
   \caption{Unsharp-masking tests. We used an \textit{F814W} ACS/WFC image of NGC~1097. From left to right and top to bottom: the resultant unsharp-masked image for a Gaussian kernel with a width of 3, 5, 10, 15, and 20 pixels respectively. The bar in the top-right corner of the first image is a reference of {physical size}, the scale is the same for all the images.  }
   \label{Figtests}
   \end{figure*}

We selected the image with the largest field of view available for each object. This makes the Advance Camera for Surveys/Wide Field Camera (ACS/WFC) the highest prioritised instrument, with a field of view of $202^{\prime \prime} \times  202^{\prime \prime}$ and a pixel scale of $0.\!\!^{\prime\prime}05$ per pixel; followed by the UV and optical channel of the Wide Field Camera 3 (WFC3/UVIS), that {covers} $160^{\prime \prime} \times  160^{\prime \prime}$, {with} $0.\!\!^{\prime\prime}04$ per pixel; and finally, although heavily featured in our work due to its extended use, the Wide Field Planetary Camera 2 (WFPC2), that uses a mosaic of 3 CCDs with a coverage of $80^{\prime \prime} \times  80^{\prime \prime}$, $0.\!\!^{\prime\prime}0996$ per pixel and a fourth smaller CCD with a coverage of $36^{\prime \prime} \times  36^{\prime \prime}$, $0.\!\!^{\prime\prime}0455$ per pixel.

Optical radiation was chosen for our work since it allows one to observe stars, dust, and star formation. Most of the selected images were taken using the \textit{F814W} filter and the rest were taken using \textit{F606W}, \textit{F555W}, and \textit{F702W}.

Our final working sample is composed of 52 galaxies with star-forming nuclear rings. Most of the galaxies in the AINUR, and hence in this work, are known to be barred. Only twelve of the galaxies in our sample are not barred.

Three of the selected images were taken with the WFC3/UVIS, thirteen using the ACS/WFC and the remaining images used were taken with the WFPC2. Five of the exposures were taken using the \textit{F606W}  filter, one was taken with \textit{F555W}, one with \textit{F702W}, and the remaining galaxies were all pictured using the \textit{F814W} filter. The complete details can be found in Table \ref{tab:tablafiltros}.

The most significant sources of incompleteness in our sample are derived from the original sample from the AINUR, including the unavailability of high-quality imaging for a substantial number of known nuclear ring hosts. The AINUR also exhibits a clear bias towards nearby galaxies due to their visibility and our resolving power, thereby precluding the study of small nuclear rings in distant galaxies.

Our complete dedication to \textit{HST} images is the primary cause of the size discrepancy between our sample and that of the AINUR. The goal of the AINUR was to account for all known nuclear rings, including those observed solely by ground-based telescopes, while our present work demands high angular resolution. Given the excluded galaxies and the limitations of the AINUR, it was concluded that our sample is not volume-complete.

 Figure~\ref{tododist} illustrates the distribution of galaxies with stellar mass, Hubble type, and bar type. The stellar masses were taken from \cite{S4G}{, when available,} and the morphology types were taken from \cite{Buta2007}{, when available,} and from NED for the rest. The family classifications SA, SAB, and SB are purely visual estimates of bar strength, but as \cite{Buta2007} point out, this classification is broadly consistent with more quantifiable bar strength measurement methods.

\section{Image processing}

\begin{figure*}[h]

 \includegraphics[height=6 cm]{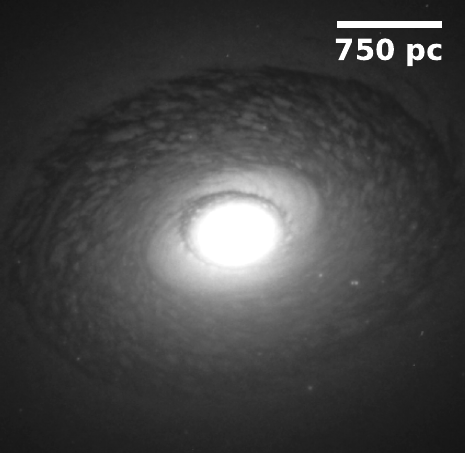}
\includegraphics[height=6 cm]{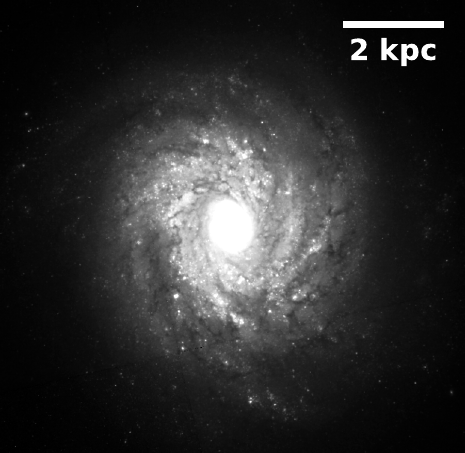}
\includegraphics[height=6 cm]{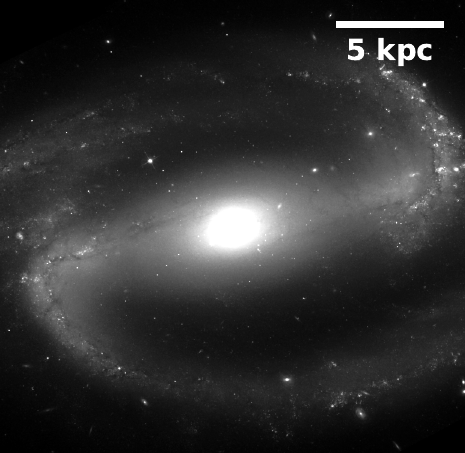}
   \caption{Morphology classification examples. From left to right, \textit{HST} images of: NGC~4459 (flocculent), NGC~3982 (multi-armed), and NGC~1300 (grand design). The instrument and filter details are in Table~\ref{tab:tablafiltros}. The bar in the top-right corner is a reference of {physical size}.}
   \label{Figmorf}
\end{figure*}

\begin{figure*}[h!]
 \includegraphics[height=6.2 cm]{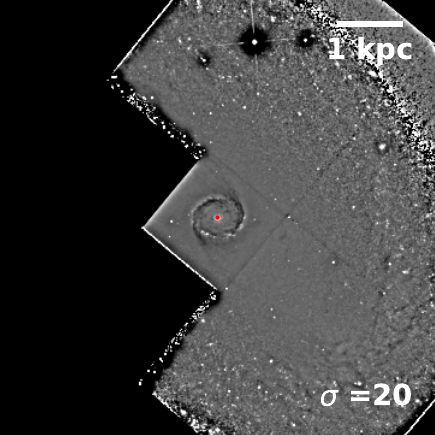}
\includegraphics[height=6.2 cm]{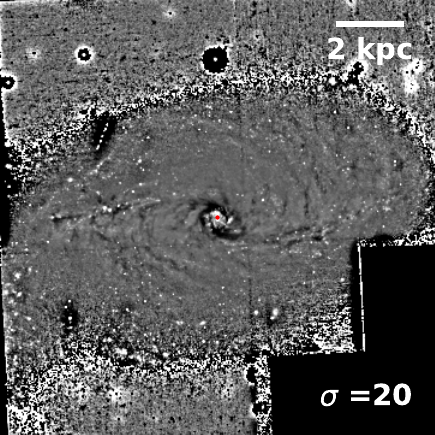}
\includegraphics[height=6.2 cm]{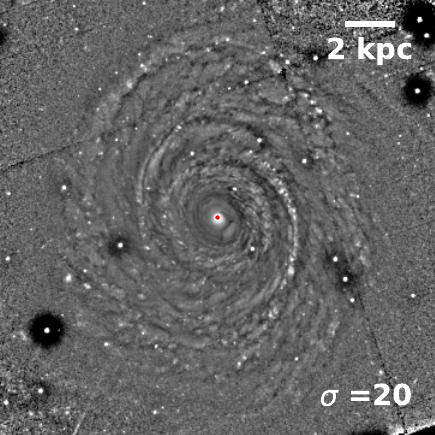}
   \caption{{Ring class examples}. From left to right, unsharp-masked images of: NGC~1512 (two-armed), NGC~7552 (twoarms+), and NGC~6753 (many-armed). The bar in the top-right corner is a reference of {physical size}. The sigma in pixels of the Gaussian kernel used to produce the unsharp-masked image appears in the bottom-right corner.}
  \label{Figring}
\end{figure*}

We use unsharp-masking to enhance the view of the inner structure of the galaxies within the selected \textit{HST} images. This technique is used to uncover { small-scale variations within the image previously obscured by the larger scale brightness distribution}. This is done by eliminating lower spatial frequencies. The resultant image reveals the dust structure in both the most external regions and closer to the centre, where it was previously obscured by the {global light distribution}. This method does not reveal any structure that was not already present in the \textit{HST} image; rather, it brings it to the forefront and facilitates its observation and characterisation. {The unsharp-masked image ($I_{\rm UM}$) is a ratio of the original image ($I_{0}$) by a convolution ($\ast$) of said original image with a Gaussian kernel ($G$):}

\begin{equation}
\centering
      I_{\rm UM} = \frac{I_{0}}{I_0 \ast G}.
   \end{equation}

This approach has been shown to produce results comparable to those achieved through the use of other methods, such as structure maps \citep*{POGGE2002}, while offering faster processing times and greater versatility, since the technique can be adjusted to better suit each case. {An early description of the digital implementation of this method can be found in \cite{SCHWEIZER1985}}. We conduct a series of tests (Fig.~\ref{Figtests}) and determine that our study requires kernels with widths ranging from 10 to 30 pixels to optimally reveal the internal dust structure of the galaxies in the sample.   

For most galaxies, we used a kernel width of 20 pixels (that corresponds to $1^{\prime\prime}$ for the ACS/WFC, $0.\!\!^{\prime\prime}8$ for the WFC3/UVIS and around $2^{\prime\prime}$ for the WFPC2), except for NGC~864, NGC~1433{,} NGC~4571, and NGC~1097 for which we used 30 pixels; and NGC~4593 and NGC~5135 for which we used 15 pixels. A method with similar basis and principles, applied to \textit{HST} data, can be found in \cite{Elme2002}.

The unsharp-masked images of the galaxies in our sample are shown in Appendix B.

\section{Sample classification}

We categorise the sample into distinct classes according to two different visual classifications. The first of these classifications is based on the morphology of the galaxy, and the second one is based on the morphology of the ring.

In order to properly classify the galaxies on the basis of their morphology, we use the \textit{Spitzer} Survey of Stellar Structure in Galaxies \citep[S$^4$G;][]{S4G} images when available. This is especially the case for galaxies for which the field of view of the \textit{HST} image is not {large} enough to cover the full extent of the galaxy.

\subsection{Galaxy arm morphology}

 \begin{figure*}
   \centering
   
   \includegraphics[height=7.81 cm]{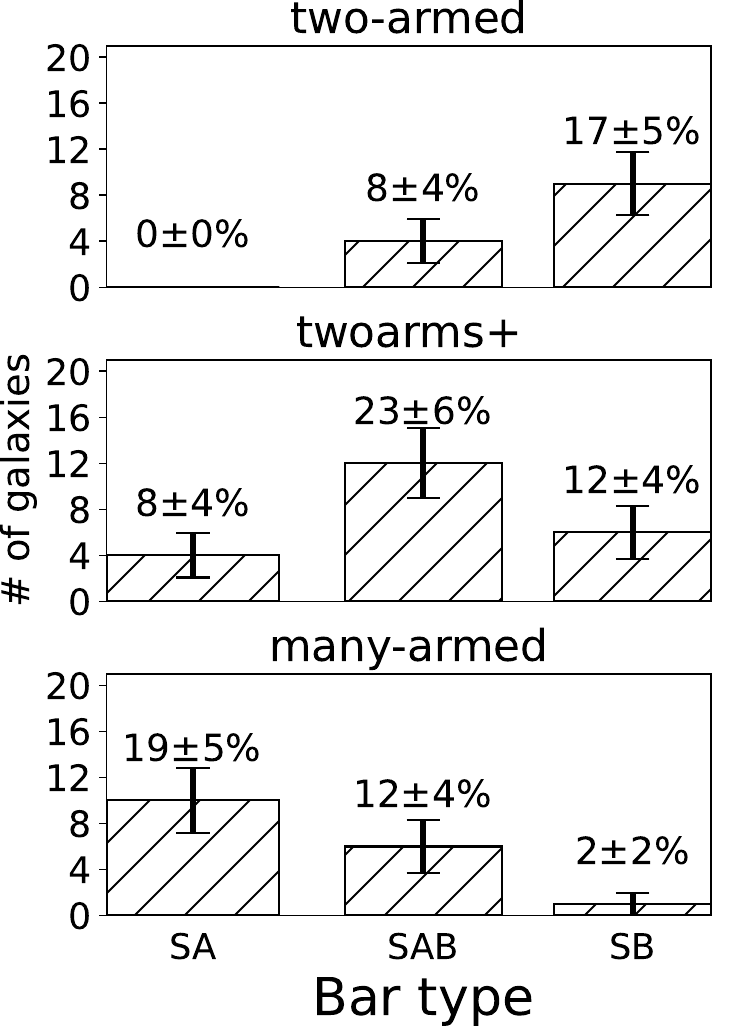}
   \includegraphics[height=7.81 cm]{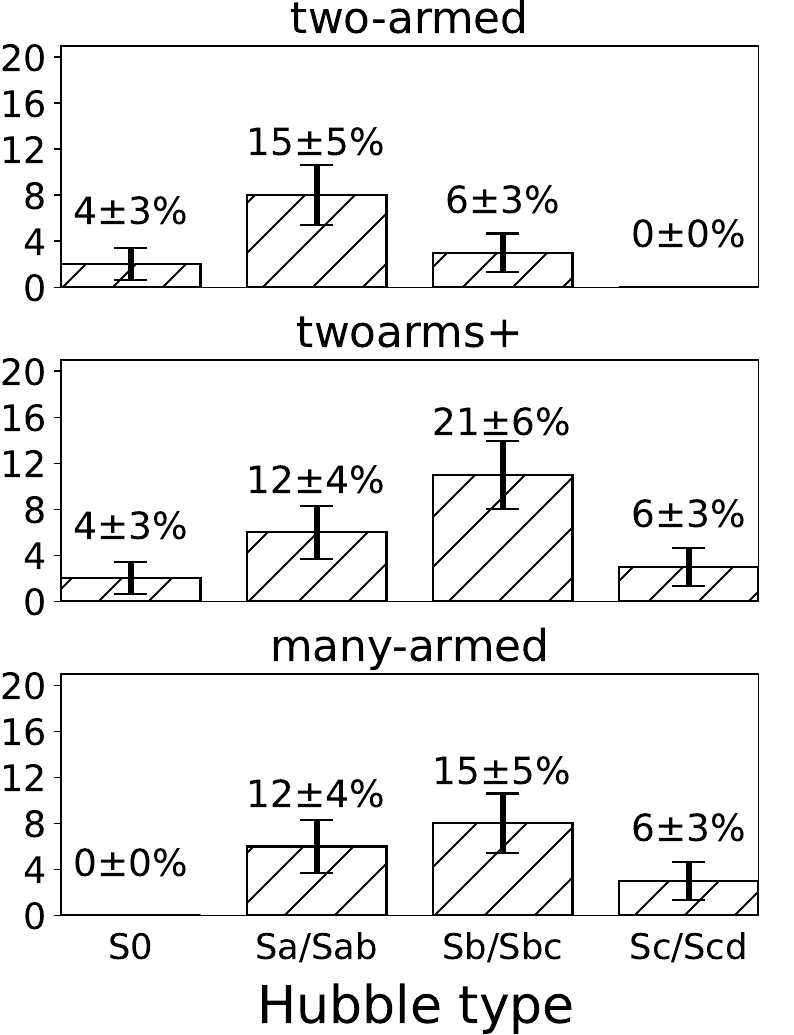}
   \includegraphics[height=7.81 cm]{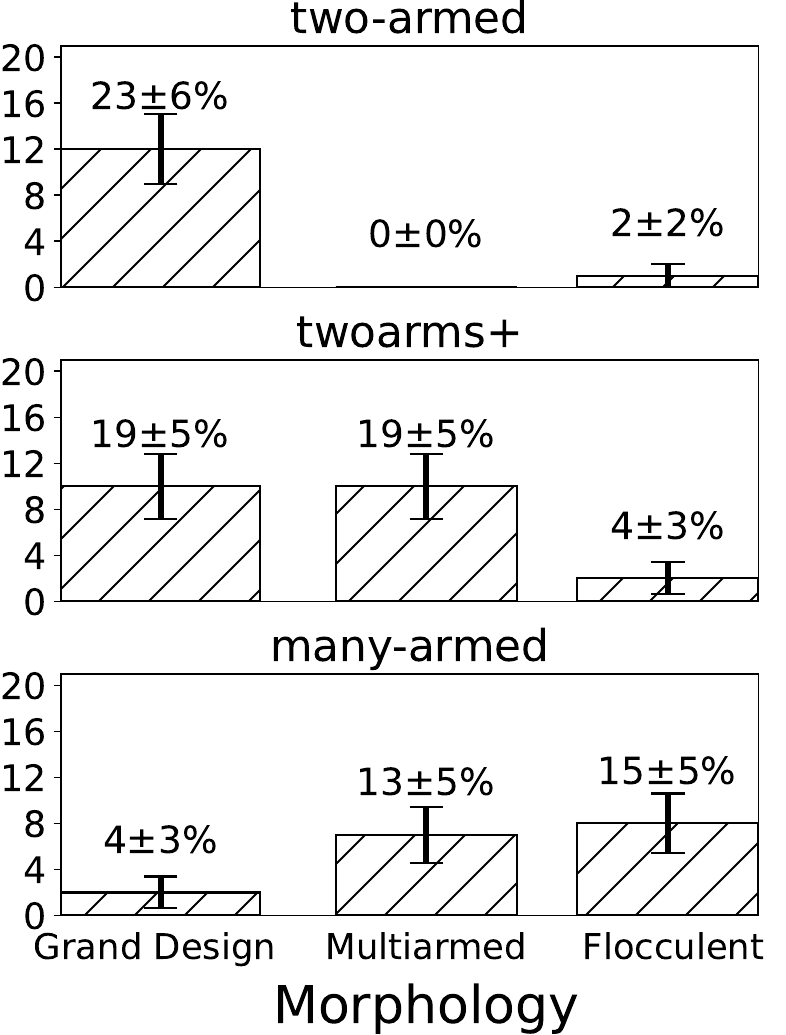}
   
   \caption{Distributions of bar type, Hubble type, and global galaxy morphology for each ring type. The percentage of galaxies belonging to each category, along with an uncertainty assigned assuming a binomial distribution, is shown for each case.}
              \label{ringdist}
    \end{figure*}

This classification is focussed on the visual structure of the galaxy as a whole, the three types are: flocculent, multi-armed and grand design galaxies. These classes are based on \cite{ELME1985}, and the classification is conducted in accordance with the parameters delineated by \cite{Elmegreen2011}.

Flocculent galaxies are characterised by a chaotic structure, devoid of any prominent arms, displaying instead a fragmented and interlaced dust spiral towards the centre accompanied by short armlets, like in NGC~4459.

Multi-armed galaxies have various long and continuous prevalent arms, like in NGC~3982. 

Finally, Grand design galaxies have a better defined structure, composed of a clear core surrounded by two well-defined arms that extend radially, like in NGC~1300.

Examples of each of these classes can be seen in Fig.~\ref{Figmorf}.

\subsection{Nuclear ring structure}
This classification focusses on the morphology of the structure that surrounds the nuclear ring, mainly the dust lanes that extend from the ring. It encompasses three distinct categories:

Two-armed: The nuclear ring is dominated by two prominent symmetric dust lanes that extend radially from the ring.

Two-armed with additional filaments (twoarms+): The nuclear ring is crowded with dust lanes, among which two stand out as the most prevalent.

Many-armed: The nuclear ring is fed by a variety of arms with no discernible hierarchical structure, like NGC~6753.

The two-armed and twoarms+ categories are both characterised by the dominance of two symmetric arms extending from the nuclear ring. The difference between these two classes lies in the presence, around the nuclear ring, of smaller armlets apart from the two main arms. For example, while the nuclear region of NGC~1512 (two-armed) is predominantly devoid of dust outside the aforementioned two main arms; NGC~7552 (twoarms+) exhibits heightened structural complexity, with the dust following pathways that deviate from the direction of the two main arms.

\section{Results}

The morphological distribution of our sample is: 24 grand design galaxies, 17 multi-armed galaxies, and 11 flocculent galaxies. The nuclear ring classification resulted in a distribution of 13 two-armed rings, 22 two-armed rings with additional filaments, and 17 many-armed rings. We cross-reference these classifications against each other and against known galaxy characteristics to find and highlight tendencies in the nuclear ring properties within our sample.

{\subsection{Results for nuclear rings}}

The bar, Hubble type and morphology distributions have been extracted for each of the previously defined ring classes (Fig.~\ref{ringdist}). We assign an uncertainty for the percentage values assuming a binomial distribution.

Two-armed rings occur exclusively in barred or weakly barred galaxies. Although twoarms+ rings may occur in unbarred galaxies, they predominantly manifest in weakly barred galaxies. The many-armed rings appear indistinctly in both unbarred and weakly barred galaxies, whereas they are very rare in strongly barred galaxies. A pattern can be seen in which each class of rings, from two-armed rings to many-armed rings, has a decreasing affinity for bars.

Two-armed rings occur mainly in earlier Hubble types, such as Sa or Sab. Twoarms+ rings and many-armed rings are much more present in later Sb or Sbc types. 

Two-armed rings are present in half of the grand design galaxies in our sample and are extremely infrequent in the other morphological classes. Almost a quarter of our sample is made up of grand design galaxies with two-armed nuclear rings. Similarly, twoarms+ rings are rare in flocculent galaxies, but are very common in grand design and multi-armed galaxies. Finally, many-armed rings avoid grand design galaxies and are observed in more than 70 $\%$ of flocculent galaxies from our sample. The overall pattern of this distribution shows a relationship between grand design galaxies and two-armed rings and another relationship between flocculent galaxies and many-armed rings.

Figure~\ref{ringdist} shows a common trend, characterised by a gradual shift in the peak of the distributions. Different global morphology characteristics are {associated with} different ring classes. The observation of congruent trends in the distributions of morphological properties can be attributed to the interdependence of these properties. According to \cite{ELME1985}, a connection exists between the size and strength of bars and the Hubble type of the host galaxy, resulting in stronger bars for earlier-type galaxies and weaker or absent bars for later-type galaxies. And according to \cite{ELME1989}, earlier type galaxies are classified as grand design more often while later type galaxies are {usually} classified as multi-armed and flocculent. The conclusion is that grand design galaxies tend to be strongly barred and earlier in type, multi-armed galaxies usually present weaker bars and are of later type, while flocculent galaxies are generally later-type and feature even weaker bars, or none at all.

We conclude that our sample of 52 galaxies with nuclear rings demonstrates the following relations between the previously discussed morphological properties of the host and the morphology of the nuclear ring itself:
\begin{itemize}
    \item Two-armed rings appear more frequently in earlier type grand design strongly barred galaxies.
    \item Twoarms+ rings are observed more often in later-type weakly barred galaxies, both grand design and multi-armed.
    \item Many-armed rings are more frequent in later type weakly barred or unbarred flocculent and multi-armed galaxies.
\end{itemize}

\begin{table}[H]     
\centering          
\caption{Poperties of each of the nuclear ring classes host galaxies.}
\begin{tabular}{c c c}
\hline 
Ring class         & $\epsilon_{\rm b}$ & $Q_{\rm g}$    \\ \hline
Two-armed  & 0.56 & 0.293 \\
Twoarms+    & 0.49 & 0.237 \\ 
Many-armed & 0.38 & 0.159 \\ \hline                           
\end{tabular}
\tablefoot{Average values of the maximum ellipticity of the bar $\epsilon_{\rm b}$ and the non-axisymmetric torque parameter $Q_{\rm g}$ \citep[from ][]{AINUR} for each ring class. }
\label{tab:tablafuerza}
\end{table}

Table~\ref{tab:tablafuerza} shows the average values taken from the AINUR of the maximum ellipticity of the bar $\epsilon_{\rm b}$ and the non-axisymmetric torque parameter $Q_{\rm g}$  \citep[which measures the impact of non-axisymmetries in the galaxy;][]{COMBESSA1981} for the galaxies in our sample that showcase each class of nuclear ring. These two parameters are a quantification of the strength of the bar: the stronger the bar, the higher $Q_{\rm g}$ and $\epsilon_{\rm b}$. Galaxies hosting two-armed rings have the strongest bars, the ones that host twoarms+ rings have slightly weaker bars, and many-armed rings appear in galaxies with very weak non-axisymmetries in general.  {Simulations by \cite{Rautiainen2004} highlight how the strength of a non-axisymmetric perturbation affects the simulated global morphology of the galaxy.}

The conclusion is that the morphology of the host galaxy and in particular the presence and strength of a bar play a fundamental role in determining the morphology of the nuclear ring and the nuclear region. {And more precisely, the strength of non-axisymmetries is a major factor in influencing the morphological characteristics of a galaxy, from its innermost to its outermost regions.} This connection may imply that different dynamical processes take part in the formation of distinct types of star-forming nuclear rings. 

\subsection{Nuclear spirals}
Nuclear spirals are described in \cite{KIM2017} as a phenomenon related to the loss of angular momentum within the inner kiloparsec and the subsequent fall of the material towards the centre. They are a network of long and narrow dust filaments that spiral inwards as they get closer to the centre. {Hydrodynamical simulations prove often their presence in low-luminosity active galactic nuclei (AGNs) and that these filaments transport matter from the nuclear region to the centre of the galaxy. However, they are rather inefficient carriers because of extensive interactions with themselves, their surrounding atmosphere, and the interstellar medium \citep{ALIG2023}.}

{Nuclear spirals} have been observed in multiple objects, some of them featured in this work, such as  NGC~1097 \citep{PRIETO2005,PRIETO2021}, NGC~6951 \citep[their Figure 2]{STORCHI2007}, ESO~428-G14 \citep[see Figure 8 of][]{PRIETO2014,MAY2018}, NGC~1566 \citep[their Figure 4]{PRIETO2021}, and M31 \citep{ALIG2023}. 

We study the regions interior to the nuclear rings following the application of the unsharp-masking technique to \textit{HST} images and, as seen in Table \ref{tab:tablafiltros}, we observe nuclear spirals in 28 galaxies ({$\sim 90\%$ of those galaxies for which the nuclear interior is resolved in the \textit{HST} image and $\sim 50\%$ of the total sample}). The remainder of galaxies in the sample show no such structure (4 galaxies) or the angular resolution and visibility of the nuclear region of the available image are inadequate to determine the presence of nuclear spirals (20 galaxies). {Although nuclear spirals have not been found inside gaseous nuclear rings in hydrodynamical models \citep{Maciejewski2004b} they} appear to be a prevalent phenomenon among galaxies in our sample whose {nuclear ring} interior has been studied, { which points to contradictions between these models and observations \citep[see e.g. discussion in][]{PRIETO2005}}. However, the limited size of this sub-sample does not permit the drawing of statistically robust conclusions.

\section{Discussion}

The kinematic origin of star-forming nuclear rings in barred galaxies has been discussed in the literature. In their study of H$\alpha$ kinematics in the barred grand design galaxy NGC~1530, \cite{ZURITA2004} improve upon the previous work of \cite{REYNAUD1998} and show how gas particles in strong bars can become entrapped in looping orbits. {This causes collisions at the far end of the bar, leading to a dynamical disturbance that propagates as shocks along the leading edge of the bar \citep[see][]{SANDERS1983}}. {These shocks trap} the gas clumps in a dust lane following an elongated orbit aligned with the bar towards the much more circular \textit{$x_2$} orbit near the nucleus and between the ILRs \citep[see][]{ENGLMAIER1997}. The shocked gas ends up accumulating in \textit{$x_2$} orbits forming a ring and, if the density gets high enough, igniting star formation. {The connection between nuclear rings and \textit{$x_2$} orbits is a known characteristic of these structures. Hydrodynamical simulations suggest that the presence of \textit{$x_2$} orbits is a requirement for the formation of nuclear rings \citep[see][]{SHETH2000,REGAN,Li_2015}.}

Two-armed rings, present in strongly barred galaxies, show signs of following this process, featuring a 180º symmetry forced by the bar potential. The two main dust lanes would correspond to the shocked gas falling towards circumnuclear \textit{$x_2$} orbits {since dust, the main component in our study, is in equilibrium with the gas \citep[see Figure 3 from][]{Li_2015}}. Furthermore, \cite{Geronbarras} find a correlation between bar strength and star formation, where stronger bars showcase more star formation in the centre, where the nuclear ring is located. This indicates that two-armed rings, the most frequent type of nuclear ring in strongly barred galaxies, accumulate material and initiate star formation more efficiently than twoarms+ and many-armed rings.

 Many-armed rings, primarily observed in unbarred galaxies, raise questions about their formation process. The absence of a bar means that their formation cannot be explained by the procedure described in \cite{ZURITA2004}. These rings are the most common in flocculent galaxies, where the uniformity of the intertwined and fragmented structure, and the resulting absence of a 180º symmetry, inhibit the formation of large dust lanes towards the centre and instead allow the appearance of many armlets. {Hydrodynamical simulations building nuclear rings usually assume a barred potential and therefore do not explore the formation processes of rings in unbarred galaxies}. However, \cite{AINUR} indicate that the nuclear ring formation mechanism in unbarred galaxies is strongly related to that in barred galaxies, quite possibly involving non-axisymmetry-induced resonances. {These non-axisymmetric features could either be different from bars, such as asymmetries induced by interactions and spiral arms, or they could be a remnant of the influence of a long-gone bar}.

Twoarms+ rings {appear more frequently in weakly barred galaxies, where \textit{$x_2$} orbits are also present \citep{Albada1982}. They} feature a weaker 180º symmetry than two-armed rings but a stronger one than many-armed rings, as seen in Table~\ref{tab:tablafuerza}. This symmetry is forced by a bar potential and has proven to be a key factor in determining the nature of the nuclear ring. Our study suggests that the different ring classes form a continuum with two-armed rings at one end and many-armed rings at the other, defined as:

\begin{itemize}
    \item Two-armed rings: 180º structure forced by a strong bar.
\item Twoarms+ rings: Weaker bars forcing a weaker 180º structure.
\item Many-armed rings: Very weak or no {bar} structure leads to the formation of many armlets.

\end{itemize}
   
\section{Conclusions}

We study 52 galaxies with nuclear rings, selected from the Atlas of Images of NUclear Rings (AINUR). We use unsharp-masked images from archival data collected by the \textit{HST} to analyse and characterise the structure of the nuclear regions of these galaxies.

We grouped the galaxies into different classes according to two visual classifications. Firstly, we categorised them based on the morphology of the galaxy as a whole, which resulted in the division of the sample into 11 flocculent galaxies, 17 multi-armed galaxies, and 24 grand design galaxies. Additionally, a classification based on the morphology of the nuclear ring itself resulted in a distribution of 13 galaxies with two-armed rings dominated by two dust lanes, 22 galaxies with twoarms+ rings that feature additional smaller armlets apart from the two main dust lanes, and 17 galaxies presenting many-armed rings with multiple similar armlets extending radially from the ring. The use of these classifications enabled the inference of correlations between the galactic properties and the structure of nuclear rings.

{We study the regions interior to the nuclear rings following the application of the unsharp-masking technique to \textit{HST} images and observe nuclear spirals in 28 galaxies from our sample ($\sim 90\%$ of those galaxies for which the interior of the nuclear ring is resolved). Although central spirals have not been found inside gaseous nuclear rings in hydrodynamical models \citep{Maciejewski2004b} they appear to be a prevalent phenomenon among galaxies in our sample whose nuclear ring interior has been studied. However, the limited size of this sub-sample does not permit the drawing of statistically robust conclusions.}

We find that two-armed rings are more common in early-type grand design barred galaxies. Twoarms+ rings are related to later-type weakly barred galaxies, both grand design and multi-armed. Lastly, many-armed rings are associated {with} later-type unbarred flocculent and multi-armed galaxies.

Galaxies hosting two-armed rings present the strongest bars. Those that host twoarms+ rings have slightly weaker bars. And many-armed rings appear in galaxies with very weak non-axisymmetries. We conclude that the global morphology of the host galaxy{, and in particular the presence and properties of a bar,} play a fundamental role in determining the morphology of the nuclear ring and the nuclear region. {Specifically, that the strength of non-axisymmetries strongly influences the morphological characteristics of a galaxy, from its outermost to its innermost regions.}

This study henceforth suggests that the different ring classes form a continuum, governed by the influence of a 180º symmetry present in the surroundings of the nuclear ring. Two-armed rings stand at one end and many-armed rings at the other, with twoarms+ rings as an intermediate case.

\begin{acknowledgements}
{We thank our anonymous referee for comments that helped to improve the paper.}

Based on observations made with the NASA/ESA \textit{Hubble Space Telescope}, and obtained from the Hubble Legacy Archive, which is a collaboration between the Space Telescope Science Institute (STScI/NASA), the Space Telescope European Coordinating Facility (ST-ECF/ESA) and the Canadian Astronomy Data Centre (CADC/NRC/CSA).

This research has made use of the NASA/IPAC Infrared Science Archive, which is funded by the National Aeronautics and Space Administration and operated by the California Institute of Technology.

This research has made use of the NASA/IPAC Extragalactic Database (NED), which is operated by the Jet Propulsion Laboratory, California Institute of Technology, under contract with the National Aeronautics and Space Administration.

SC acknowledges funding from the State Research Agency (AEI-MICIU) of the Spanish Ministry of Science, Innovation, and Universities under the grant ‘The relic galaxy NGC 1277 as a key to understanding massive galaxies at cosmic noon’ with reference PID2023-149139NB-I00.

AP acknowledges J. Beckman and  P. Inwin for discussions, and spanish project PID2023-146851NB-100.

Co-funded by the European Union (MSCA Doctoral Network EDUCADO, GA 101119830 and Widening Participation, ExGal-Twin, GA 101158446). JHK acknowledges grant PID2022-136505NB-I00 funded by MCIN/AEI/10.13039/501100011033 and EU, ERDF.

\end{acknowledgements}

\bibliography{MIREF}

\clearpage
\onecolumn
\begin{appendix}
\section{Sample table}
\setcounter{table}{0}
\renewcommand{\thetable}{A.\arabic{table}}
\begin{table*}[h!]
\centering
\caption{Properties of the nuclear ring host galaxies and of the images used in our work.}
\resizebox{6.25 in}{!}{
\begin{tabular}{lllllllll}
\hline
ID         & Dist. & Morph. type     & Instrument & Filter  & Morph.  & Ring class & Nuc. &$\log(M_*$ \\ 
    &  (Mpc) &   &   & & class & & spiral  &$/{\,M_\odot})$ \\
 (1)   &  (2) &  (3) & (4)  & (5) & (6) & (7) & (8)  &(9) \\ \hline
ESO~565-11 & 66.5           & (R$^\prime$)SB(rs)a     & WFPC2      & \textit{F814W}    & GD           & two-armed    & \texttimes   &-               \\ \hline
IC~342     & 3.3            & SAB(rs)cd       & WFPC2      & \textit{F814W}    & M   & twoarms+   & ?       &-              \\ \hline
NGC~278  & 11.7           & SAB(rs)b        & WFPC2      & \textit{\textit{F814W} }   & F   & many-armed & ?    &-                \\ \hline
NGC~613  & 18.7           & SB(rs)bc        & WFPC2      & \textit{F814W}    & GD     & twoarms+   & ?      &11.09          \\ \hline
NGC~864    & 21.8           & SAB(rs)c        & WFPC2      & \textit{F814W}    & M   & twoarms+   & ?    &10.18          \\ \hline
NGC~1068 & 15.3           & (R)SA(rs)b      & ACS/WFC    & \textit{\textit{F814W} }    & M   & twoarms+   & \checked     &10.86          \\ \hline
NGC~1097   & 15.2           & (R$^\prime$)SB(rs)b pec  & ACS/WFC    & \textit{F814W}    & GD           & twoarms+  & \checked     &11.24           \\ \hline
NGC~1241 & 55.8           & SAB(rs)b        & WFPC2      & \textit{F606W}    & M   & twoarms+   & ?     &-                 \\ \hline
NGC~1300   & 20.2           & SB(rs)b         & ACS/WFC    & \textit{F814W}    & GD           & two-armed    & \checked    &10.58          \\ \hline
NGC~1317   & 23.9           & (R$^\prime$)SAB(rl)a    & WFPC2      & \textit{F814W}    & F   & two-armed    & \checked   &-                 \\ \hline
NGC~1326   & 16.1           & (R$_1$)SAB(l)0/a   & WFPC2      & \textit{F814W}    & GD           & two-armed    & \checked    &10.55          \\ \hline
NGC~1433   & 11.6           & (R$^\prime$)SB(r)ab     & WFC3/UVIS  & \textit{F814W}    & GD           & two-armed    & \checked     &10.30          \\ \hline
NGC~1512   & 9.5            & (R$^\prime$)SB(r)ab pec & WFPC2      & \textit{F814W}    & GD           & two-armed    & \checked     &10.33          \\ \hline
NGC~1566   & 17.4           & (R$^\prime$)SAB(s)bc    & ACS/WFC    & \textit{F814W}    & GD           & two-armed    & \checked     &10.58          \\ \hline
NGC~1672   & 15.0           & (R:)SB(r)bc     & ACS/WFC    & \textit{F814W}    & GD           & twoarms+   & \checked     &10.66         \\ \hline
NGC~1808   & 10.9           & (R$_1$)SAB(s)b pec  & ACS/WFC    & \textit{F814W}    & F   & many-armed & ?     &10.61          \\ \hline
NGC~2903   & 8.9            & SAB(rs)bc       & ACS/WFC    & \textit{F814W}    & GD           & many-armed & ?     &10.66          \\ \hline
NGC~2985   & 22.6           & (R$^\prime$)SA(rs)ab    & ACS/WFC    & \textit{F814W}    & F   & many-armed  & ?    &10.86          \\ \hline
NGC~2997   & 13.1           & SAB(s)c         & WFPC2      & \textit{F814W}    & M   & twoarms+   & \checked    &-                \\ \hline
NGC~3081   & 31.8           & (RR)SAB(r)0/a   & WFPC2      & \textit{F814W}    & GD           & twoarms+   & \checked     &-               \\ \hline
NGC~3185   & 18.9           & (R)SB(r)a       & WFPC2      & \textit{F814W}    & GD           & twoarms+ & ?    &10.21          \\ \hline
NGC~3310   & 17.4           & SA(rs)bc pec    & WFPC2      & \textit{F814W}    & M   & twoarms+   & ?    &10.319          \\ \hline
NGC~3504   & 23.8           & (R$^\prime$)SAB(rs)ab   & WFPC2      & \textit{F606W}    & GD           & twoarms+  & \checked    &10.40          \\ \hline
NGC~3982   & 23.6           & SAB(r)b         & WFPC2      & \textit{F814W}    & M   & many-armed & \checked      &10.26         \\ \hline
NGC~4100 {(!)}  & 18.6           & (R$^\prime$)SA(s)bc     & ACS/WFC    & \textit{F814W}    & M   & many-armed & ?     &10.58          \\ \hline
NGC~4102   & 18.6           & SAB(s)b?        & WFPC2      & \textit{F814W}    & M   & many-armed & ?    &10.554          \\ \hline
NGC~4245   & 15.0           & SB(r)0/a        & WFPC2      & \textit{F814W}    & GD           & two-armed    & \texttimes    &9.80           \\ \hline
NGC~4274   & 15.6           & (R$^\prime$)SB(r)ab     & WFPC2      & \textit{F555W}    & F   & many-armed & ?     &10.76          \\ \hline
NGC~4303   & 23.1           & SAB(rs)bc       & WFPC2      & \textit{F814W}    & M   & twoarms+   & \checked     &10.86          \\ \hline
NGC~4314   & 16.4           & (R$^\prime$)SB(r$^\prime$l)a    & WFPC2      & \textit{F814W}    & GD           & two-armed    & \checked     &10.14           \\ \hline
NGC~4321   & 24.0           & SAB(s)bc        & WFPC2      & \textit{F702W}    & GD           & twoarms+   & \checked     &10.93         \\ \hline
NGC~4448   & 11.7           & (R)SB(r)ab      & WFPC2      & \textit{F814W}    & F   & twoarms+   & ?      &10.848          \\ \hline
NGC~4459   & 16.1           & SA(r)0$^{+}$         & WFC3/UVIS  & \textit{F814W}    & F   & twoarms+  & \texttimes     &-               \\ \hline
NGC~4571   & 16.8           & SA(r)c          & WFPC2      & \textit{F814W}    & M   & many-armed & ?     &10.254          \\ \hline
NGC~4593   & 35.3           & (R$^\prime$)SB(rs)ab    & WFPC2      & \textit{F606W}    & GD           & two-armed    & \checked     &10.93          \\ \hline
NGC~4736   & 5.2            & (R)SAB(rs)ab    & WFPC2      & \textit{F814W}    & GD           & many-armed & \checked     &10.52         \\ \hline
NGC~4800   & 14.5           & SA(rs)b         & ACS/WFC    & \textit{F814W}    & M   & many-armed & \checked     &10.46         \\ \hline
NGC~4826 {(!)}  & 7.5            & (R$^\prime$)SA(r)ab pec & WFPC2      & \textit{F814W}    & F   & many-armed & ?     &10.44          \\ \hline
NGC~5033  {(!)} & 15.4           & SA(s)c          & WFPC2      & \textit{F814W}    & M   & many-armed & \checked     &10.95          \\ \hline
NGC~5135   & 57.4           & SB(l)ab         & WFPC2      & \textit{F606W}    & GD           & two-armed    & ?      &-               \\ \hline
NGC~5248   & 17.9           & (R$^\prime$)SAB(rs)bc   & WFPC2      & \textit{F814W}    & GD           & twoarms+   & \checked     &10.67          \\ \hline
NGC~5427   & 39.4           & SA(rs)bc        & WFPC2      & \textit{F606W}    & M   & twoarms+   & ?    &10.737          \\ \hline
NGC~5728   & 39.7           & (R$_1$)SAB(r)a     & WFC3/UVIS  & \textit{F814W}    & GD           & twoarms+   & \checked     &10.85          \\ \hline
NGC~5806   & 20.5           & SAB(s)b         & ACS/WFC    & \textit{F814W}    & M   & twoarms+   & \checked    &10.58         \\ \hline
NGC~6503   & 5.2            & SA(s)cd         & ACS/WFC    & \textit{F814W}    & F   & many-armed  & ?     &9.70          \\ \hline
NGC~6753   & 41.8           & (R)SA(r)b       & WFPC2      & \textit{F814W}    & M   & many-armed & \checked     &-               \\ \hline
NGC~6782   & 52.5           & (RR)SB(r)a      & WFPC2      & \textit{F814W}    & GD           & two-armed    & \checked     &-               \\ \hline
NGC~6951   & 24.4           & SAB(rs)bc       & WFPC2      & \textit{F814W}    & GD           & two-armed    & \checked     &-               \\ \hline
NGC~7217   & 16.0           & (R)SA(r)ab      & WFPC2      & \textit{F814W}    & F   & many-armed  & \texttimes    &-               \\ \hline
NGC~7469   & 70.7           & (R$^\prime$)SAB(rs)a    & ACS/WFC    & \textit{F814W}    & M   & twoarms+   & \checked     &-               \\ \hline
NGC~7552   & 20.2           & (R$^\prime$)SB(s)ab     & WFPC2      & \textit{F814W}    & GD           & twoarms+   & ?     &10.52          \\ \hline
NGC~7742   & 24.2           & SA(r)ab         & WFPC2      & \textit{F814W}    & F   & many-armed & \checked     &10.34          \\ \hline
\end{tabular}}
\tablefoot{Identification (col.~1), distance in Mpc (col.~2, from \cite{AINUR}), morphological type (col.~3, from \cite{Buta2007} when available and from NED for the rest), instrument and filter of the analysed image (col.~4 and col.~5), morphology class (col.~6, GD for grand design, M for multi-armed and F for flocculent), nuclear ring class (col.~7), nuclear spiral (col.~8, \checked ~ for galaxies that exhibit a nuclear spiral, \texttimes ~ for those that do not, and ? if it is uncertain) and stellar mass (col.~9, from \cite{S4G} when available). A sign {(!)} appears next to very tilted galaxies.}

\label{tab:tablafiltros}
\end{table*}

\clearpage
\section{Unsharp-masked images}
Unsharp-masked images obtained from the \textit{HST} images of galaxies in our sample. The galaxy ID appears in the top-left corner. At the left of each pair of images, we present an image whose size is twice the diameter of the nuclear ring. At the right we present an image whose size is five times the diameter of the nuclear ring.  The sigma in pixels of the Gaussian kernel used to produce the unsharp-masked image appears in the bottom-right corner. The bar in the top-right corner is a reference of {physical size}. The red ellipse indicates the nuclear ring present in the galaxy \citep[from][]{AINUR}. North is up and east is left.

\begin{figure*}[h]
\centering
 \includegraphics[width=.24\linewidth]{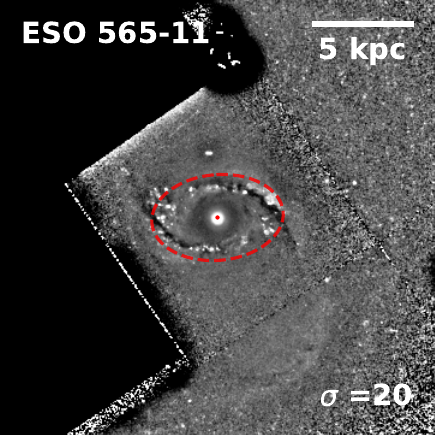}
\includegraphics[width=.24\linewidth]{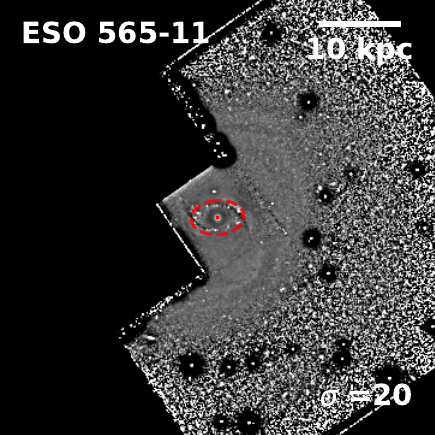}
\hspace{0.8 pt}
 \includegraphics[width=.24\linewidth]{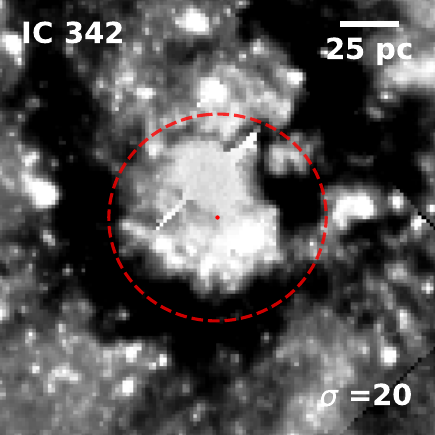}
\includegraphics[width=.24\linewidth]{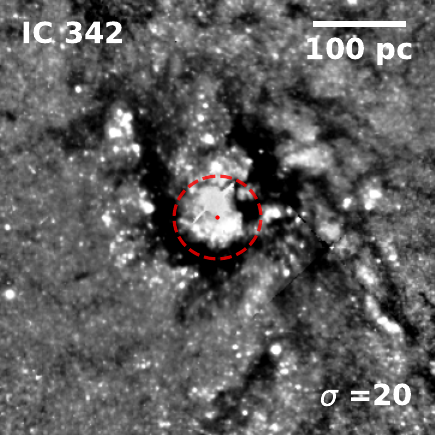}\\ 
\vspace{4.5 pt}
 \includegraphics[width=.24\linewidth]{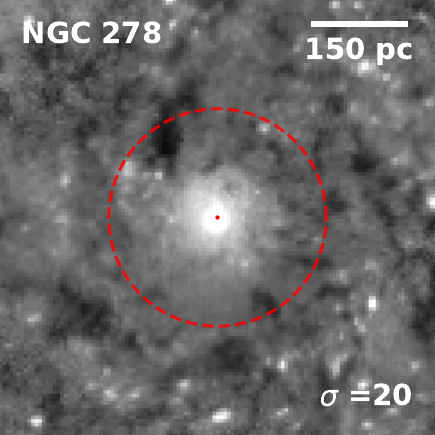}
\includegraphics[width=.24\linewidth]{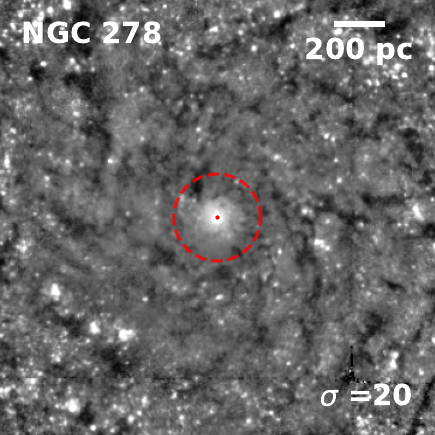}
\hspace{0.8 pt}
 \includegraphics[width=.24\linewidth]{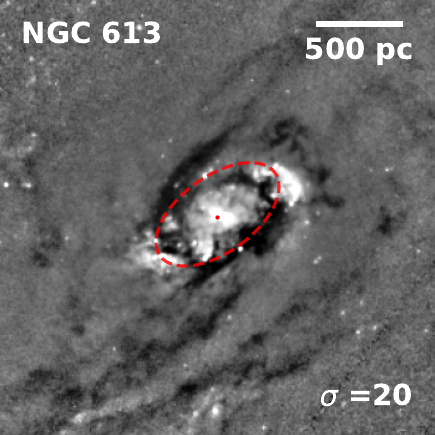}
\includegraphics[width=.24\linewidth]{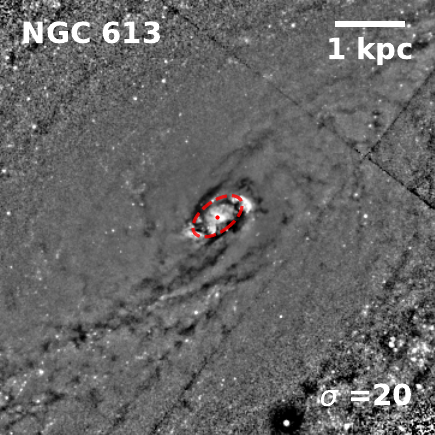}\\ 
\vspace{4.5 pt}
 \includegraphics[width=.24\linewidth]{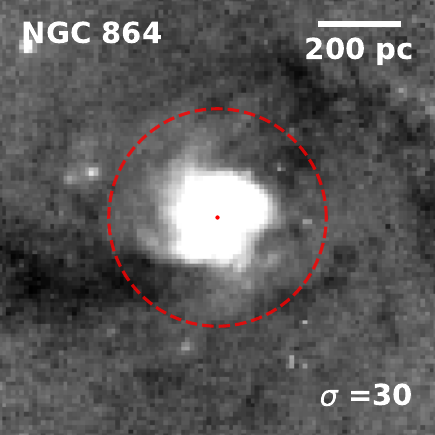}
\includegraphics[width=.24\linewidth]{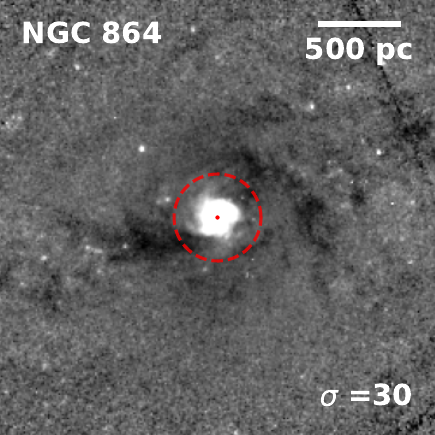}
\hspace{0.8 pt}
 \includegraphics[width=.24\linewidth]{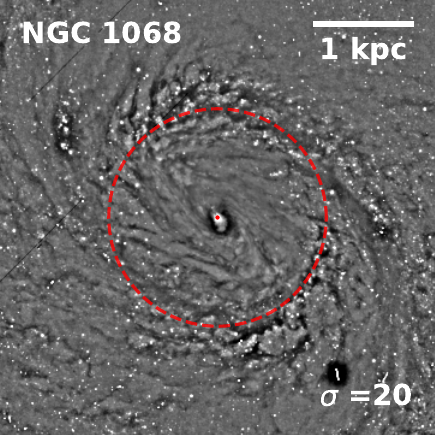}
\includegraphics[width=.24\linewidth]{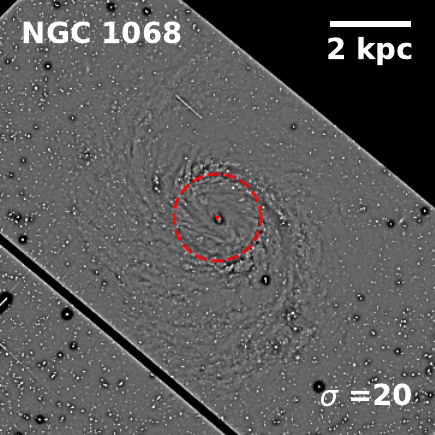}\\ \vspace{4.5 pt}
 \includegraphics[width=.24\linewidth]{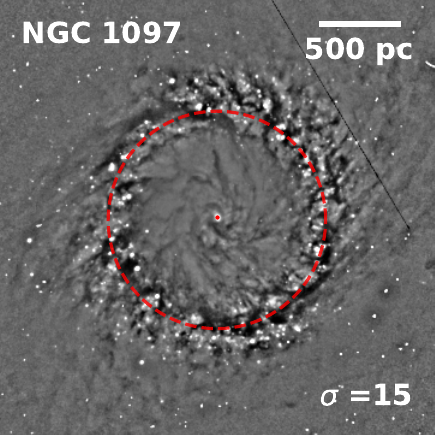}
\includegraphics[width=.24\linewidth]{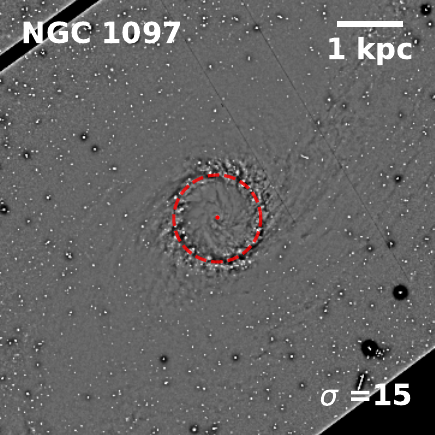}
\hspace{0.8 pt}
 \includegraphics[width=.24\linewidth]{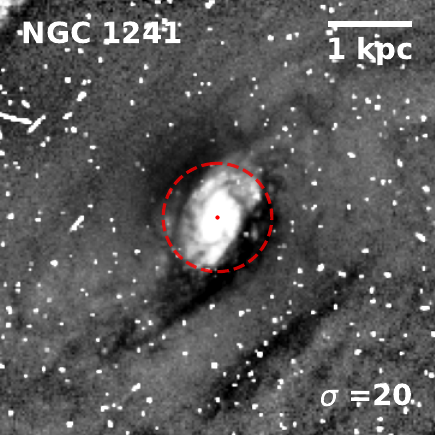}
\includegraphics[width=.24\linewidth]{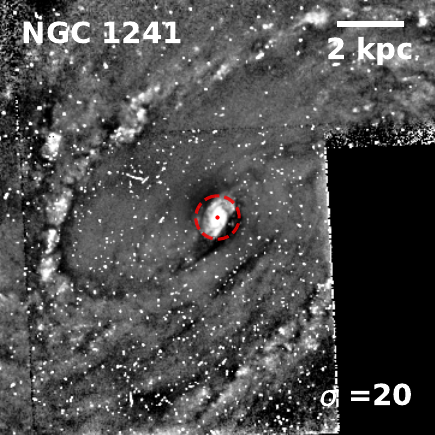}\\ \vspace{4.5 pt}
\caption{Unsharp-masked images.}
\end{figure*}
\begin{figure*}[h]
\ContinuedFloat
\centering
 \includegraphics[width=.24\linewidth]{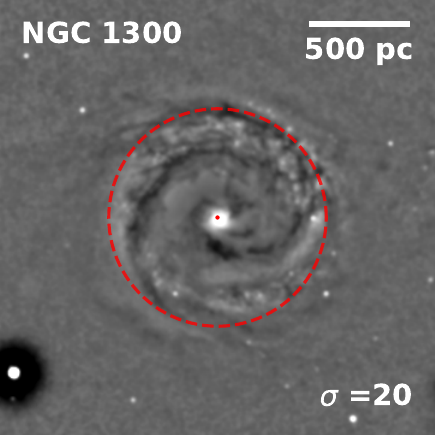}
\includegraphics[width=.24\linewidth]{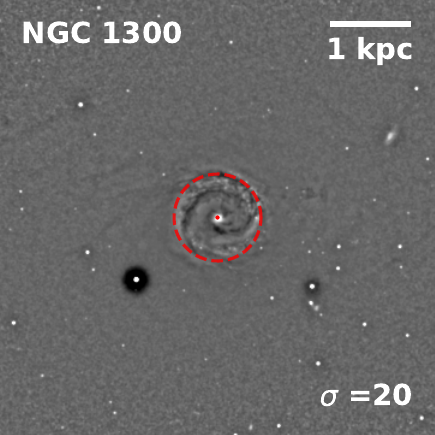}
\hspace{0.8 pt}
 \includegraphics[width=.24\linewidth]{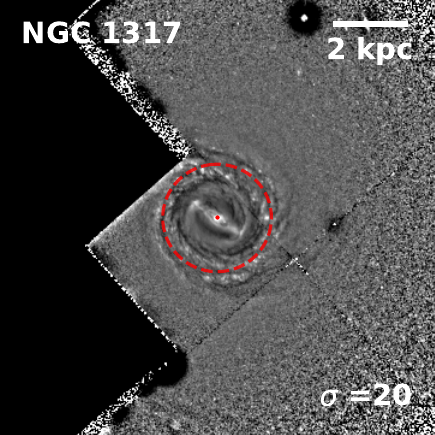}
\includegraphics[width=.24\linewidth]{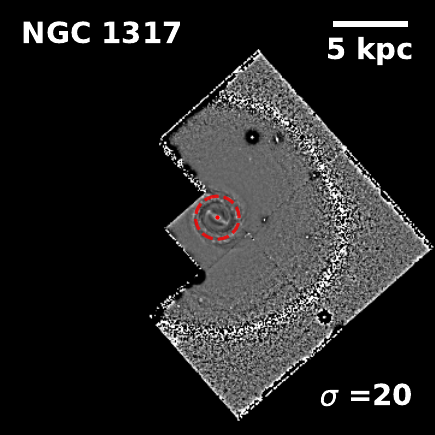}\\ \vspace{4.5 pt}

 \includegraphics[width=.24\linewidth]{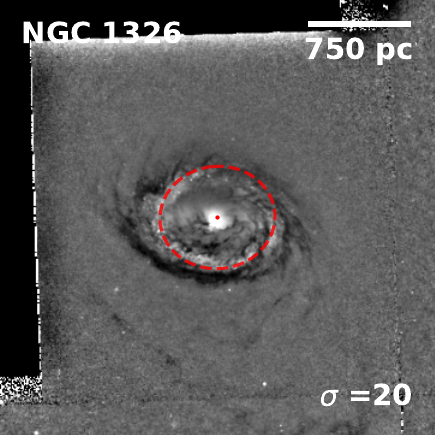}
\includegraphics[width=.24\linewidth]{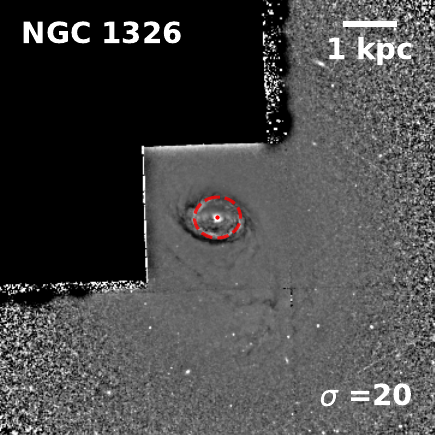}
\hspace{0.8 pt}
 \includegraphics[width=.24\linewidth]{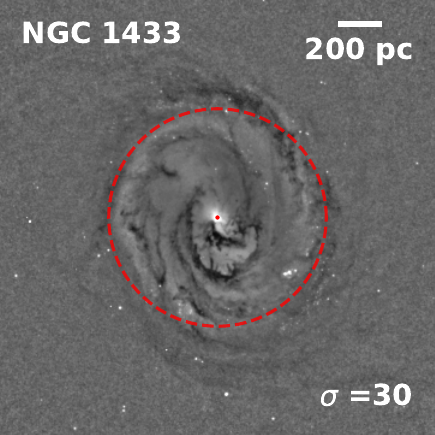}
\includegraphics[width=.24\linewidth]{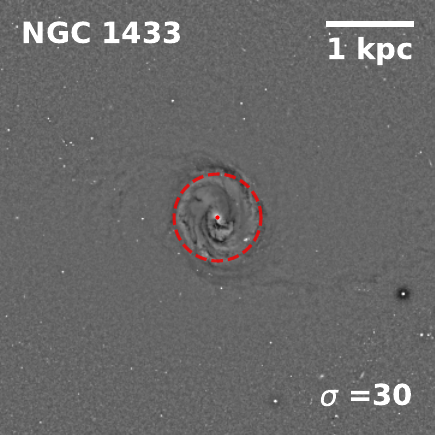}\\ 
\vspace{4.5 pt}
 \includegraphics[width=.24\linewidth]{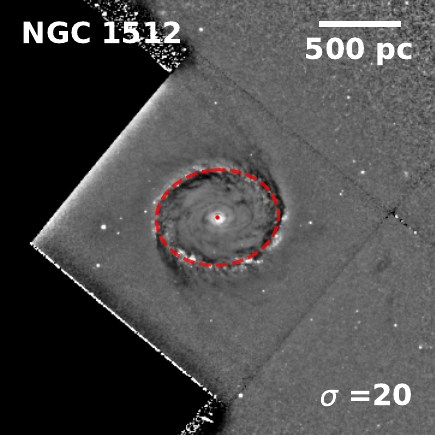}
\includegraphics[width=.24\linewidth]{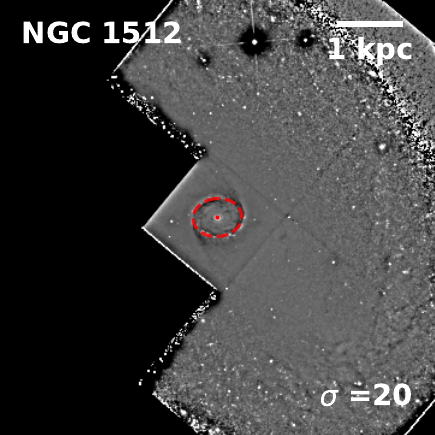}
\hspace{0.8 pt}
 \includegraphics[width=.24\linewidth]{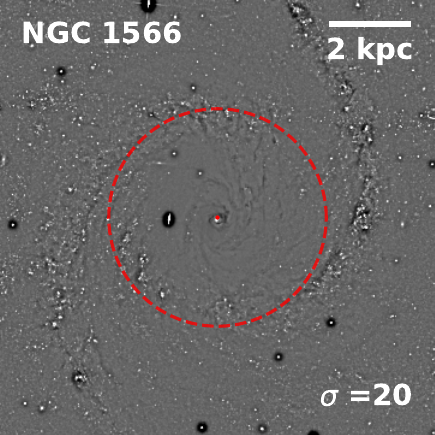}
\includegraphics[width=.24\linewidth]{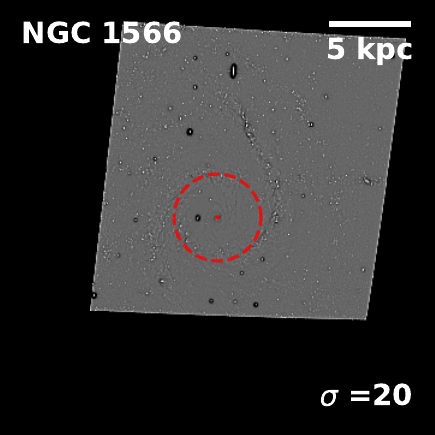}\\ 
\vspace{4.5 pt}
 \includegraphics[width=.24\linewidth]{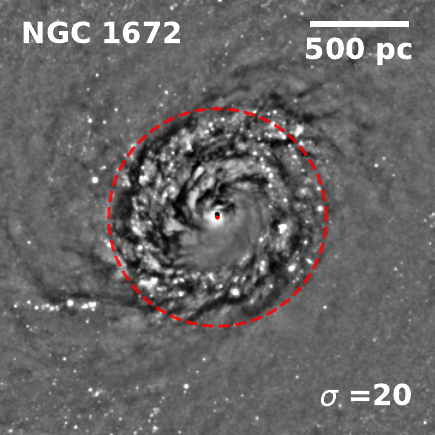}
\includegraphics[width=.24\linewidth]{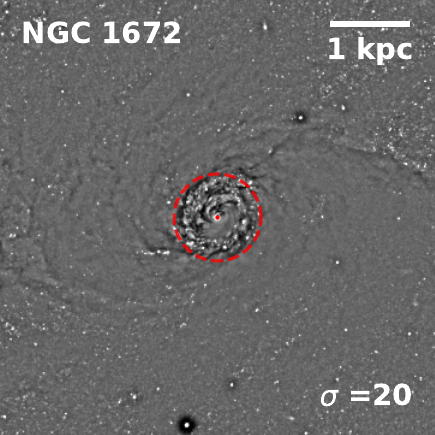}
\hspace{0.8 pt}
 \includegraphics[width=.24\linewidth]{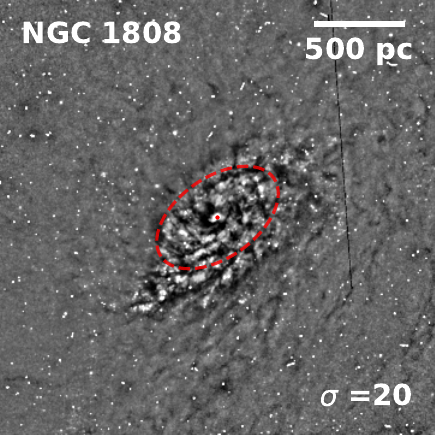}
\includegraphics[width=.24\linewidth]{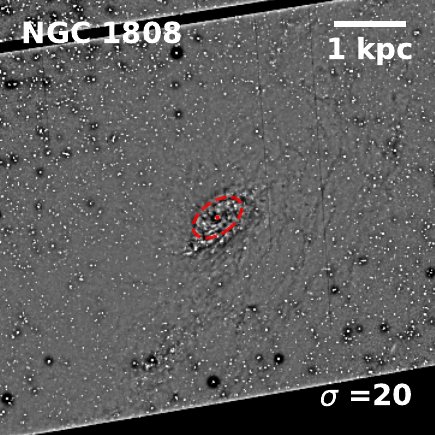}\\ 
\vspace{4.5 pt}
 \includegraphics[width=.24\linewidth]{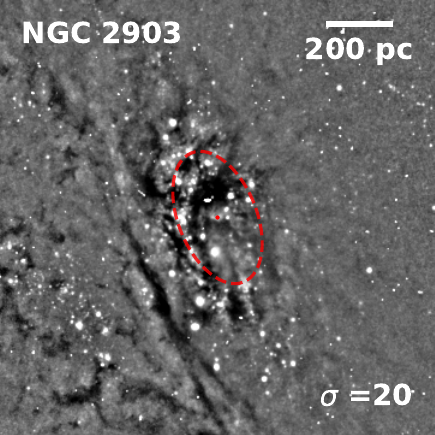}
\includegraphics[width=.24\linewidth]{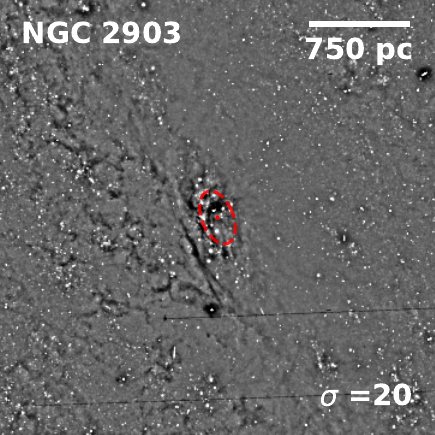}
\hspace{0.8 pt}
 \includegraphics[width=.24\linewidth]{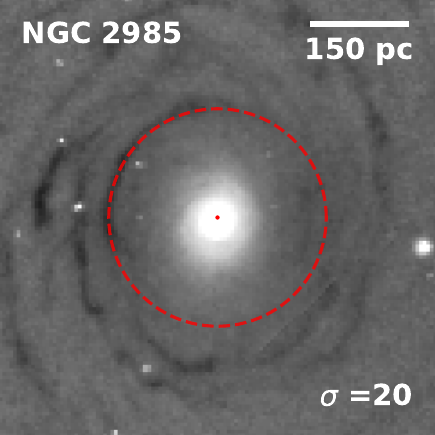}
\includegraphics[width=.24\linewidth]{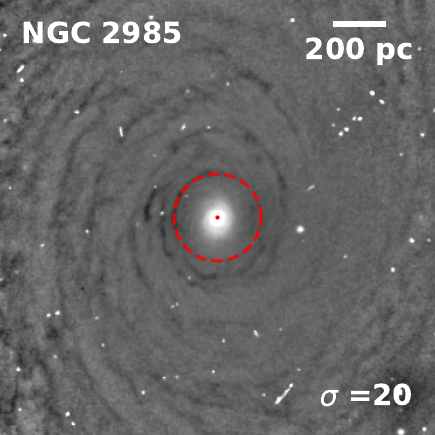}\\ 
\vspace{4.5 pt}
\caption{Continued.}
\end{figure*}

\begin{figure*}[h]
\ContinuedFloat
\centering
 \includegraphics[width=.24\linewidth]{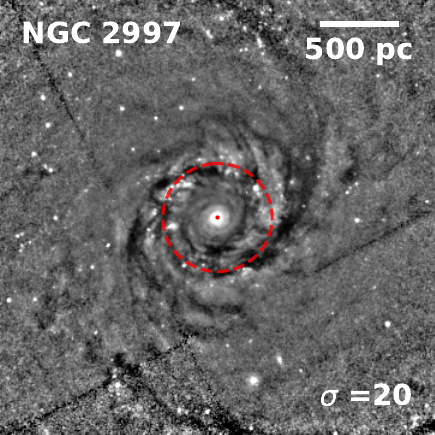}
\includegraphics[width=.24\linewidth]{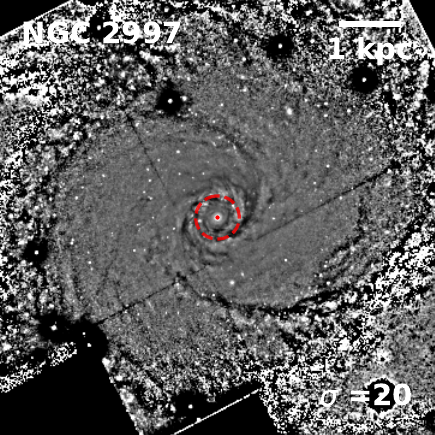}
\hspace{0.8 pt}
 \includegraphics[width=.24\linewidth]{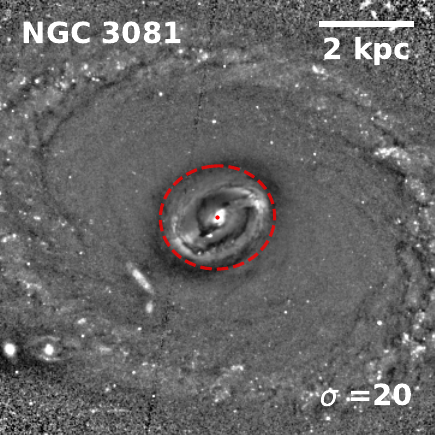}
\includegraphics[width=.24\linewidth]{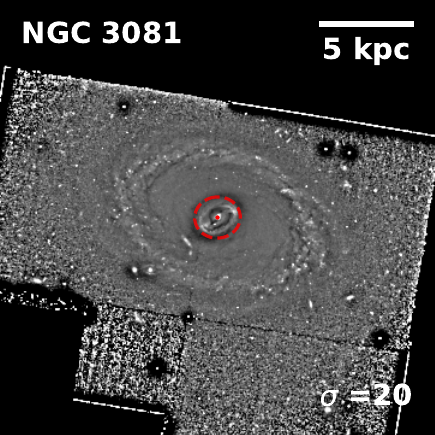}\\ 
\vspace{4.5 pt}
 \includegraphics[width=.24\linewidth]{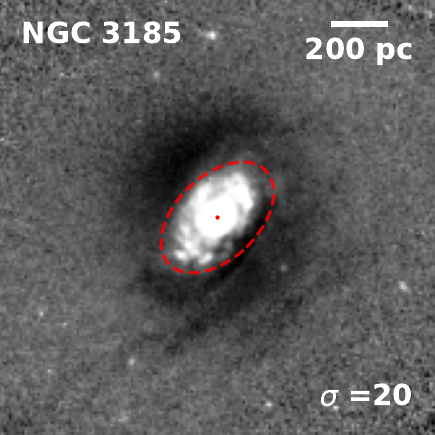}
\includegraphics[width=.24\linewidth]{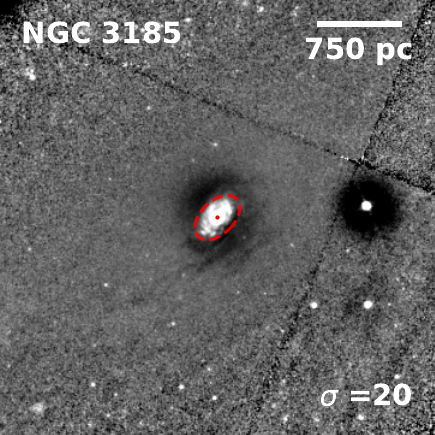}
\hspace{0.8 pt}
 \includegraphics[width=.24\linewidth]{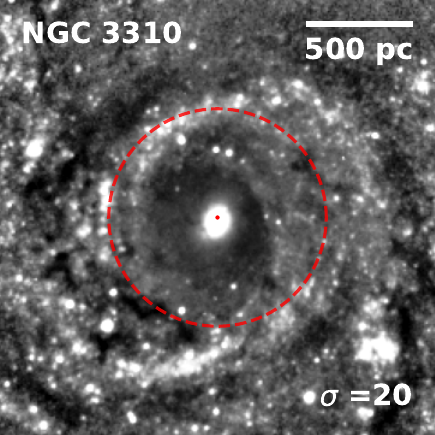}
\includegraphics[width=.24\linewidth]{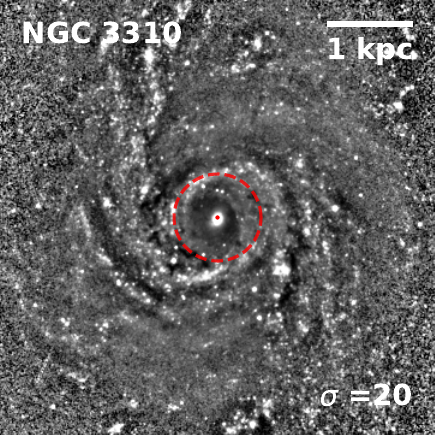}\\ 
\vspace{4.5 pt}
 \includegraphics[width=.24\linewidth]{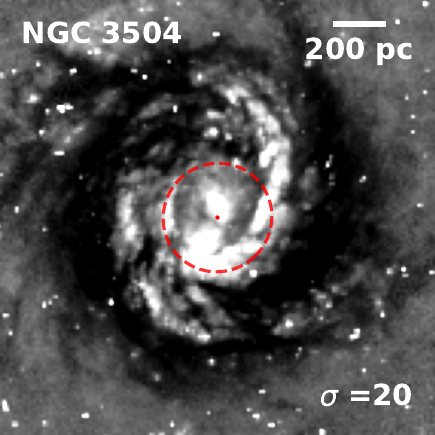}
\includegraphics[width=.24\linewidth]{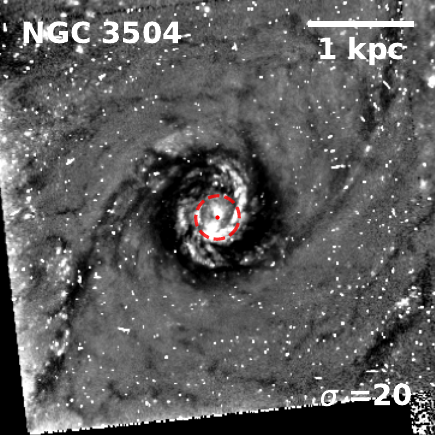}
\hspace{0.8 pt}
 \includegraphics[width=.24\linewidth]{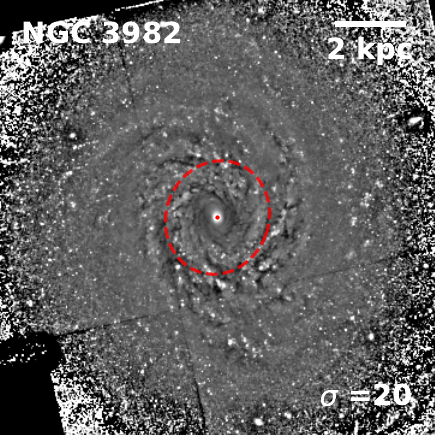}
\includegraphics[width=.24\linewidth]{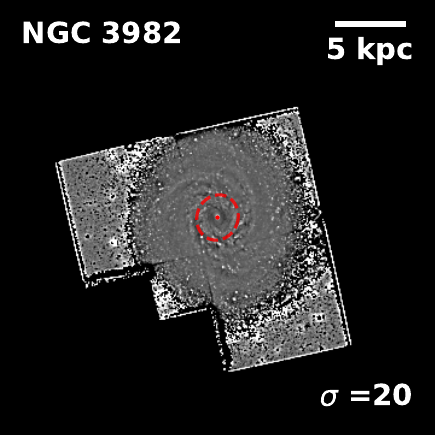}\\ 
\vspace{4.5 pt}
 \includegraphics[width=.24\linewidth]{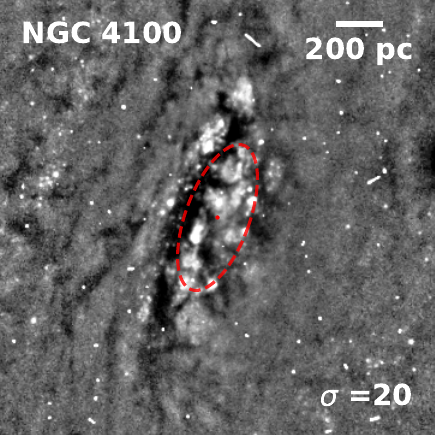}
\includegraphics[width=.24\linewidth]{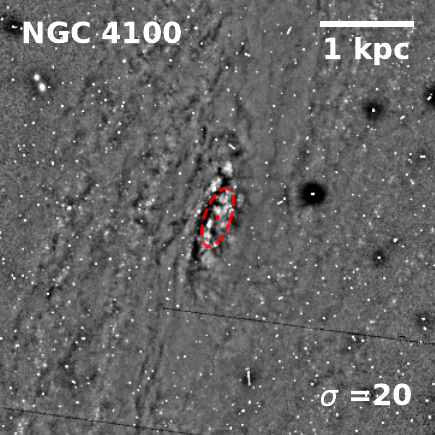}
\hspace{0.8 pt}
 \includegraphics[width=.24\linewidth]{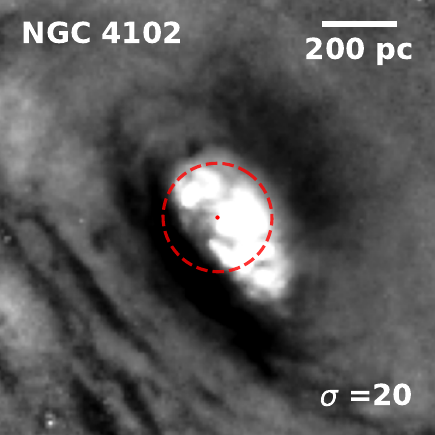}
\includegraphics[width=.24\linewidth]{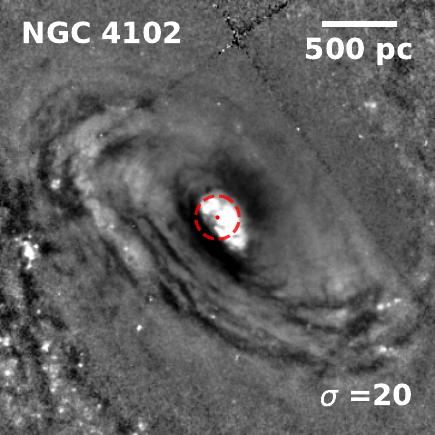}\\ 
\vspace{4.5 pt}
 \includegraphics[width=.24\linewidth]{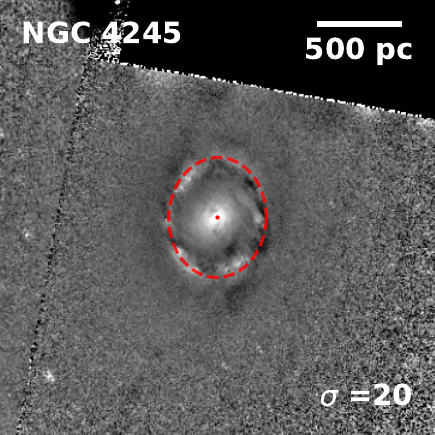}
\includegraphics[width=.24\linewidth]{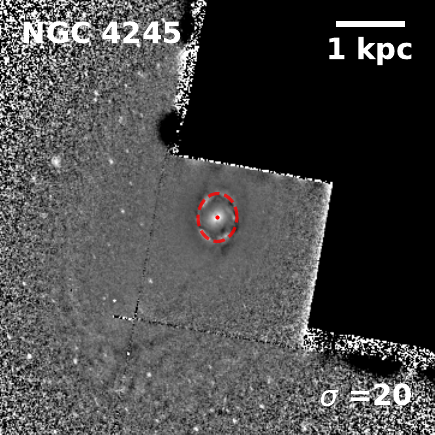}
\hspace{0.8 pt}
 \includegraphics[width=.24\linewidth]{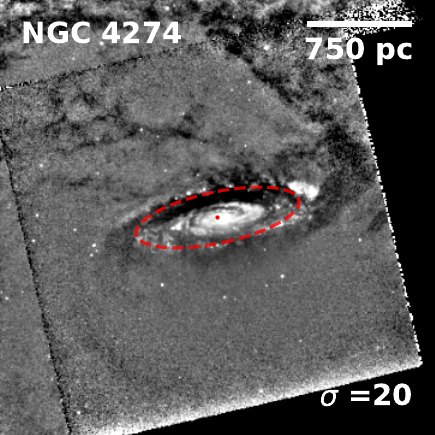}
\includegraphics[width=.24\linewidth]{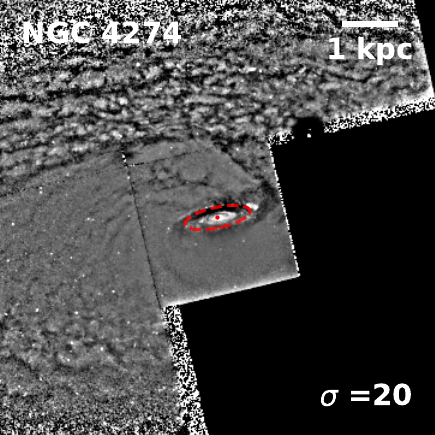}\\ 
\vspace{4.5 pt}
\caption{Continued.}
\end{figure*}

\begin{figure*}[h]
\ContinuedFloat
\centering
 \includegraphics[width=.24\linewidth]{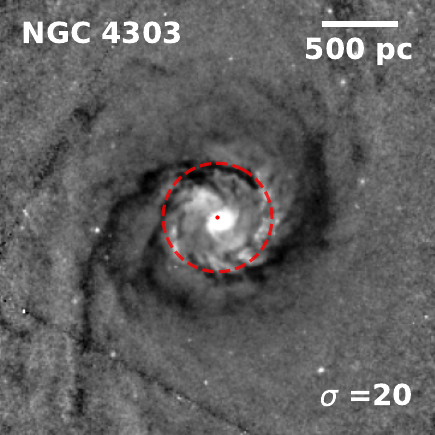}
\includegraphics[width=.24\linewidth]{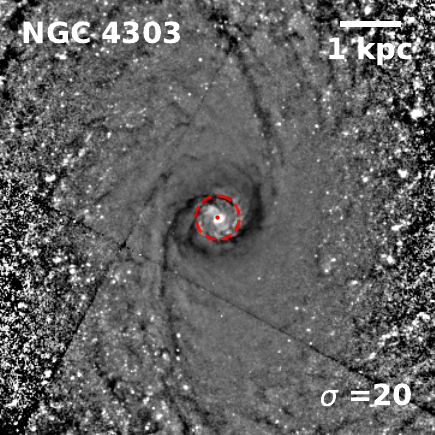}
\hspace{0.8 pt}
 \includegraphics[width=.24\linewidth]{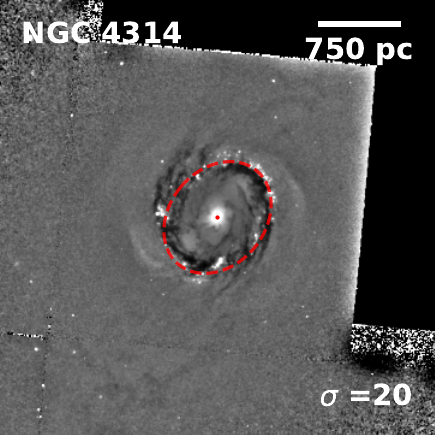}
\includegraphics[width=.24\linewidth]{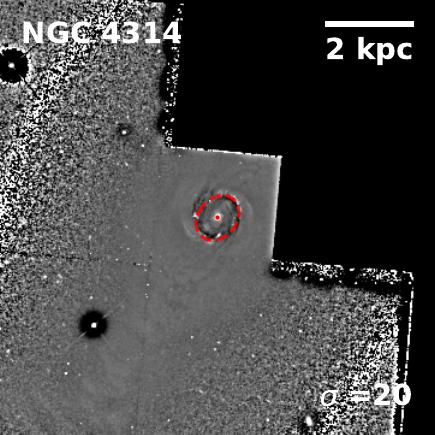}\\ 
\vspace{4.5 pt}
 \includegraphics[width=.24\linewidth]{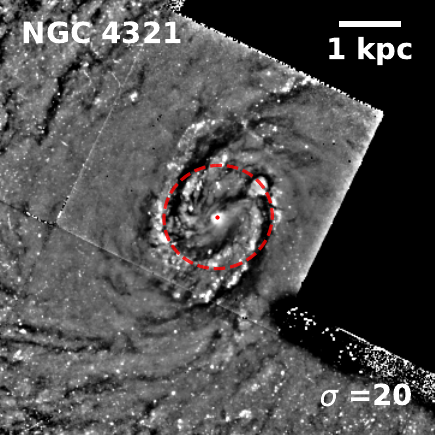}
\includegraphics[width=.24\linewidth]{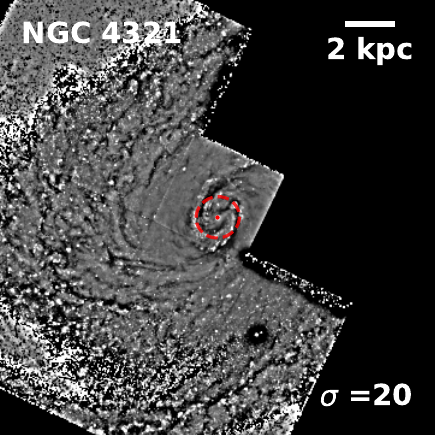}
\hspace{0.8 pt}
 \includegraphics[width=.24\linewidth]{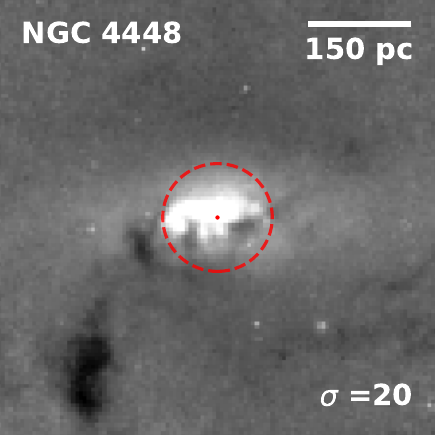}
\includegraphics[width=.24\linewidth]{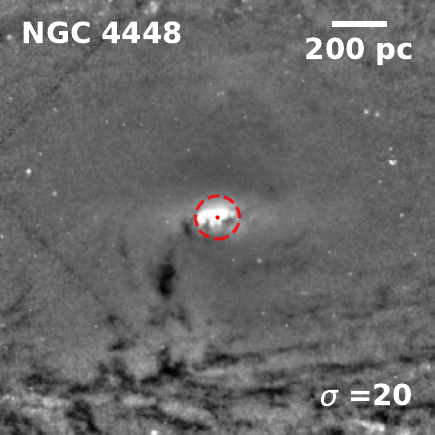}\\ 
\vspace{4.5 pt}
 \includegraphics[width=.24\linewidth]{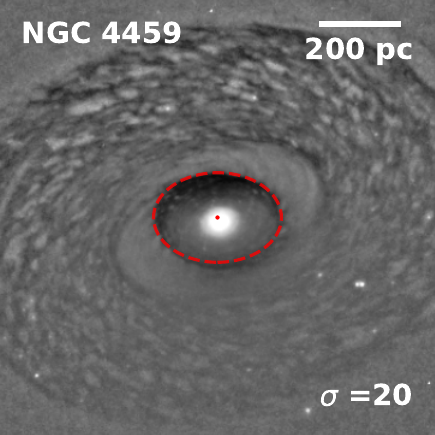}
\includegraphics[width=.24\linewidth]{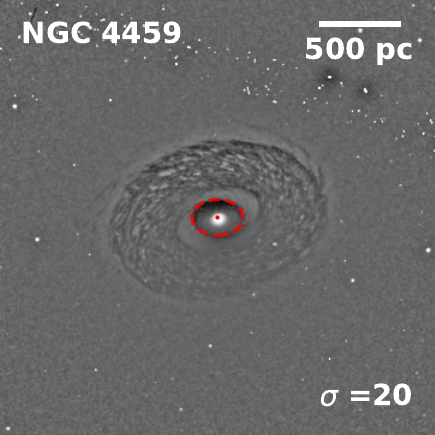}
\hspace{0.8 pt}
 \includegraphics[width=.24\linewidth]{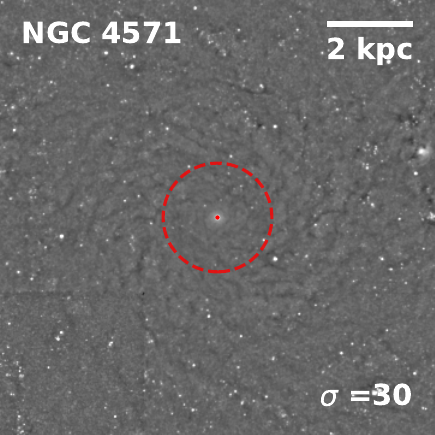}
\includegraphics[width=.24\linewidth]{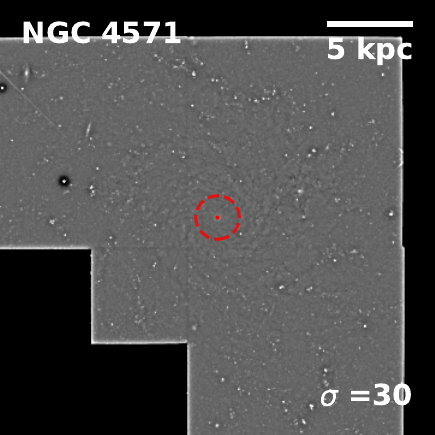}\\ 
\vspace{4.5 pt}
 \includegraphics[width=.24\linewidth]{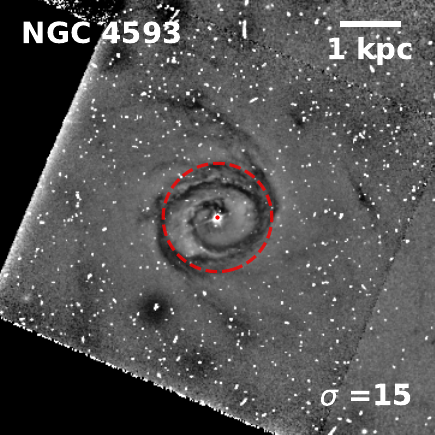}
\includegraphics[width=.24\linewidth]{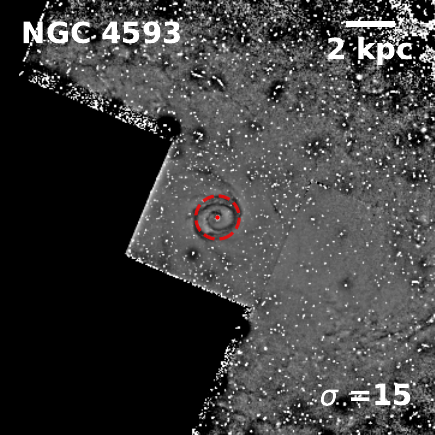}
\hspace{0.8 pt}
 \includegraphics[width=.24\linewidth]{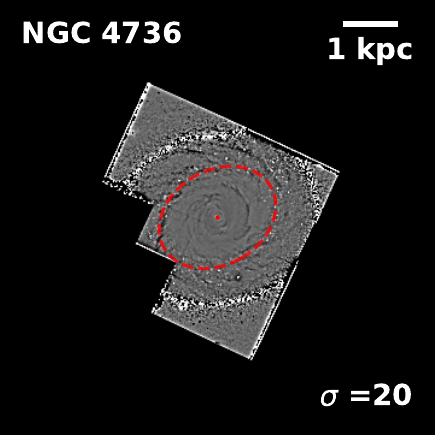}
\includegraphics[width=.24\linewidth]{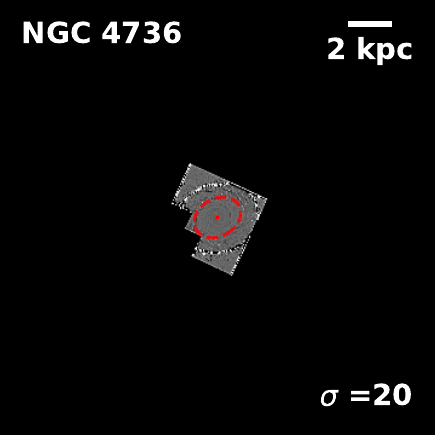}\\ 
\vspace{4.5 pt}
 \includegraphics[width=.24\linewidth]{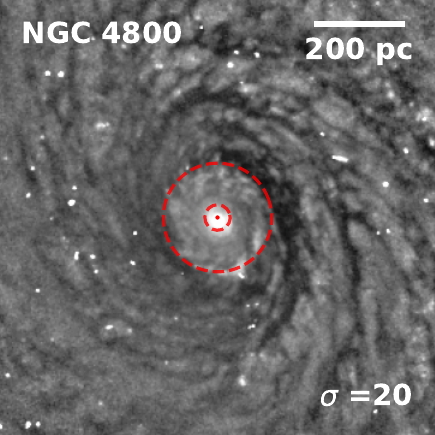}
\includegraphics[width=.24\linewidth]{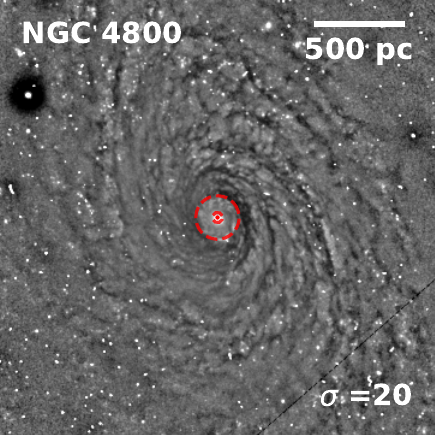}
\hspace{0.8 pt}
 \includegraphics[width=.24\linewidth]{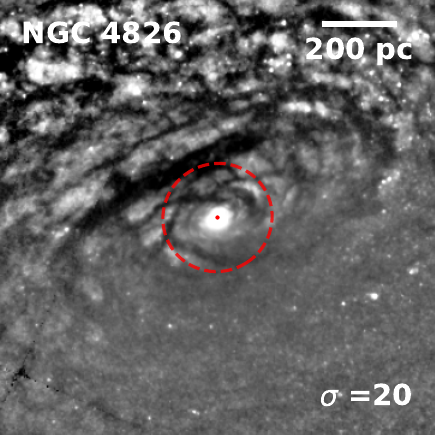}
\includegraphics[width=.24\linewidth]{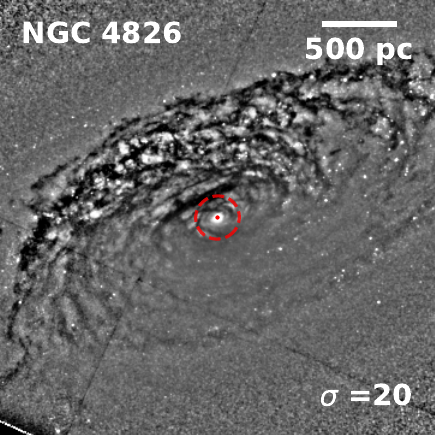}\\ 
\vspace{4.5 pt}
\caption{Continued.}
\end{figure*}

\begin{figure*}[h]
\ContinuedFloat
\centering
 \includegraphics[width=.24\linewidth]{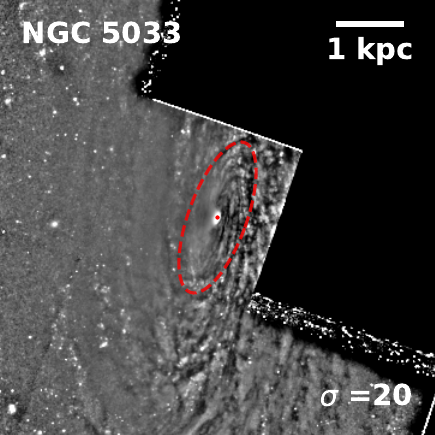}
\includegraphics[width=.24\linewidth]{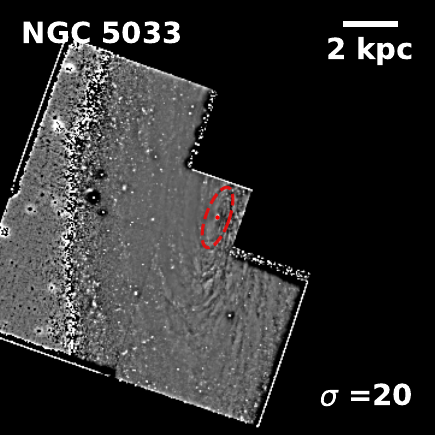}
\hspace{0.8 pt}
 \includegraphics[width=.24\linewidth]{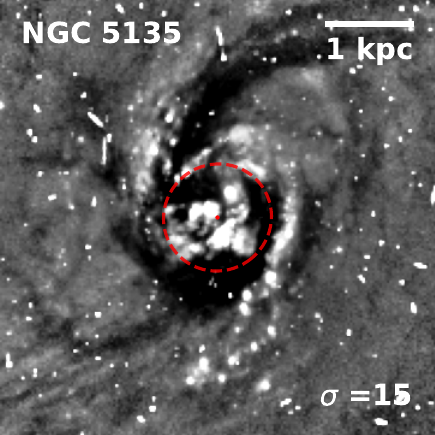}
\includegraphics[width=.24\linewidth]{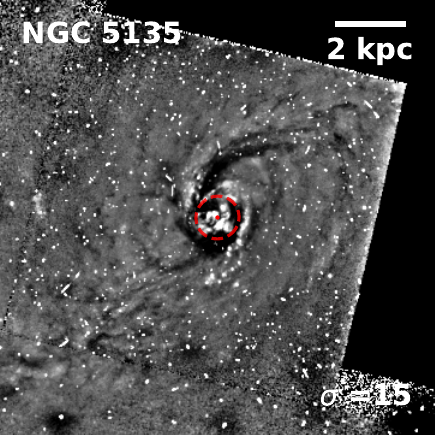}\\ 
\vspace{4.5 pt}
 \includegraphics[width=.24\linewidth]{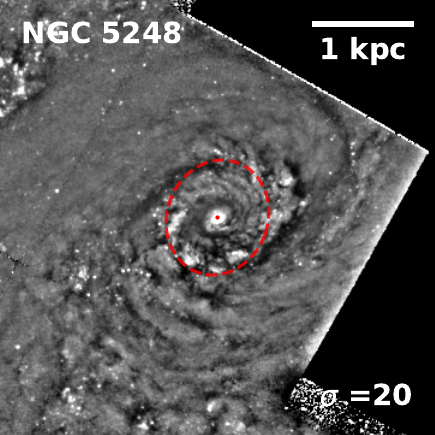}
\includegraphics[width=.24\linewidth]{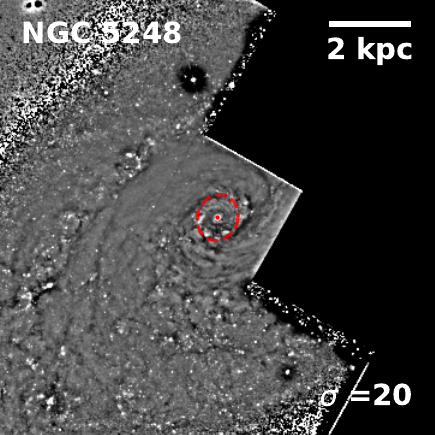}
\hspace{0.8 pt}
 \includegraphics[width=.24\linewidth]{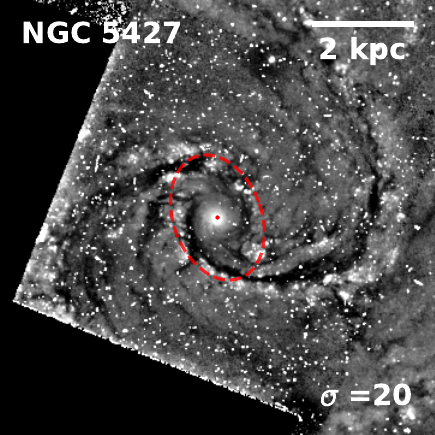}
\includegraphics[width=.24\linewidth]{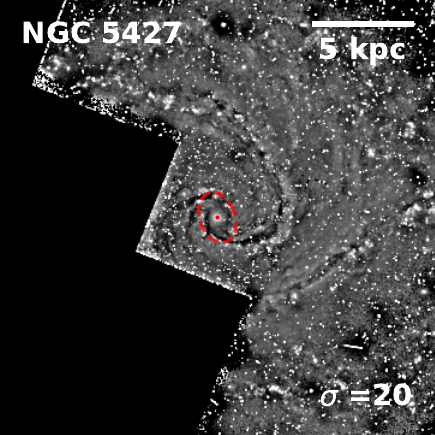}\\ 
\vspace{4.5 pt}
 \includegraphics[width=.24\linewidth]{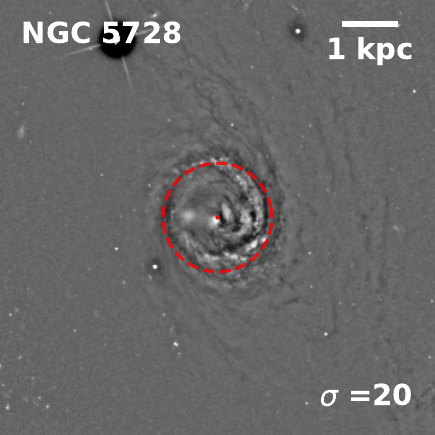}
\includegraphics[width=.24\linewidth]{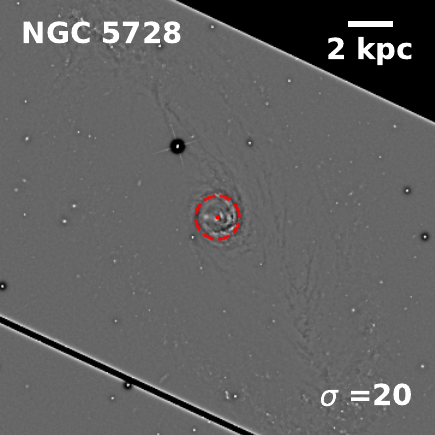}
\hspace{0.8 pt}
 \includegraphics[width=.24\linewidth]{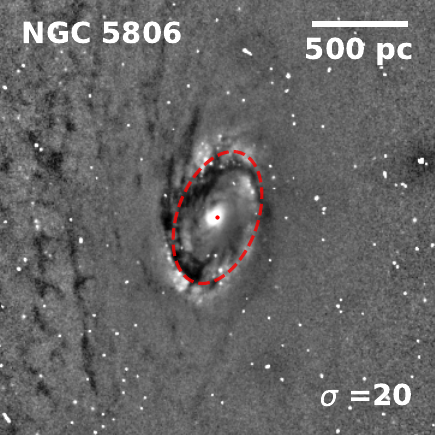}
\includegraphics[width=.24\linewidth]{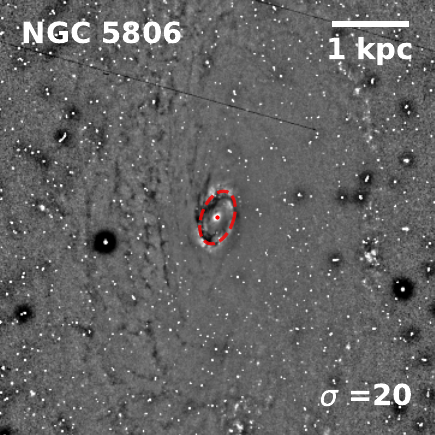}\\ 
\vspace{4.5 pt}
 \includegraphics[width=.24\linewidth]{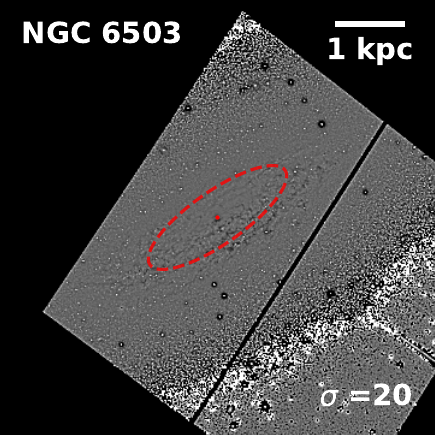}
\includegraphics[width=.24\linewidth]{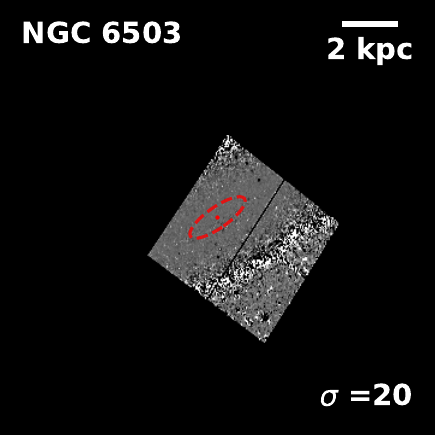}
\hspace{0.8 pt}
 \includegraphics[width=.24\linewidth]{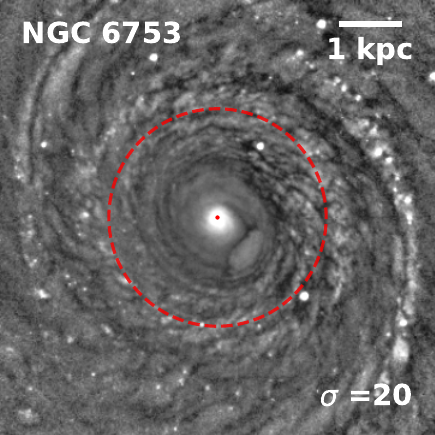}
\includegraphics[width=.24\linewidth]{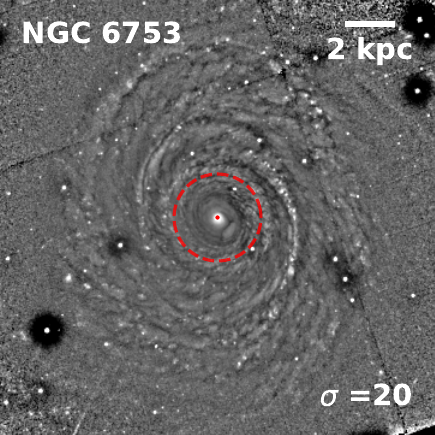}\\ 
\vspace{4.5 pt}
 \includegraphics[width=.24\linewidth]{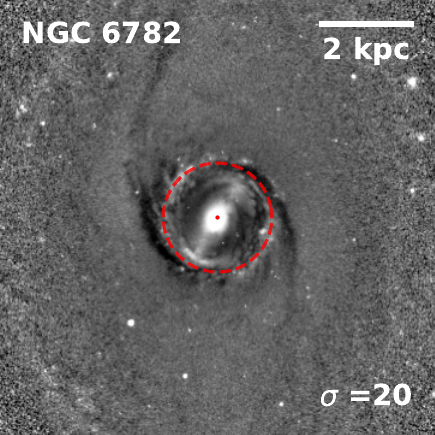}
\includegraphics[width=.24\linewidth]{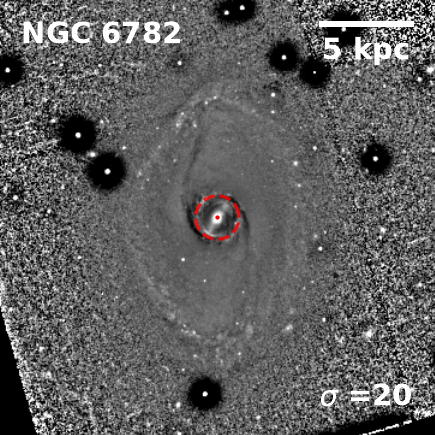}
\hspace{0.8 pt}
 \includegraphics[width=.24\linewidth]{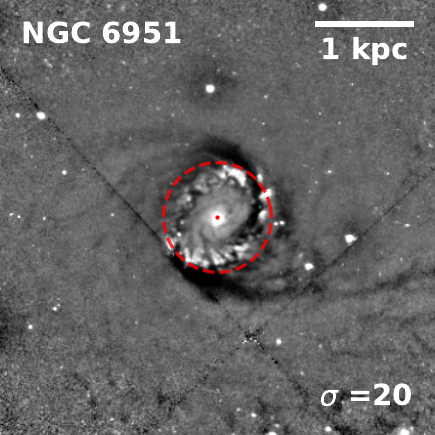}
\includegraphics[width=.24\linewidth]{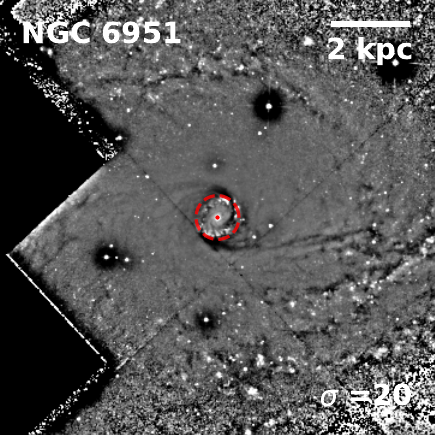}\\ 
\vspace{4.5 pt}
\caption{Continued.}
\end{figure*}

\begin{figure*}[h]
\ContinuedFloat
\centering
 \includegraphics[width=.24\linewidth]{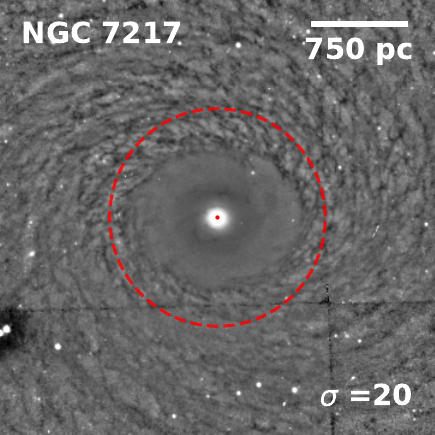}
\includegraphics[width=.24\linewidth]{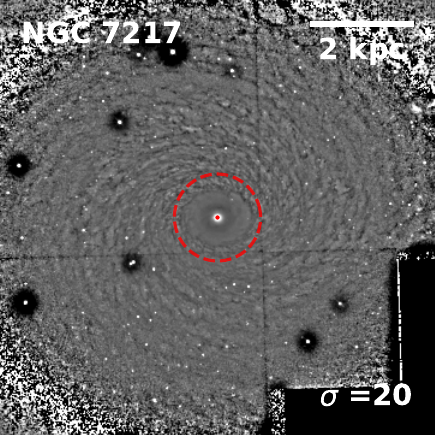}
\hspace{0.8 pt}
 \includegraphics[width=.24\linewidth]{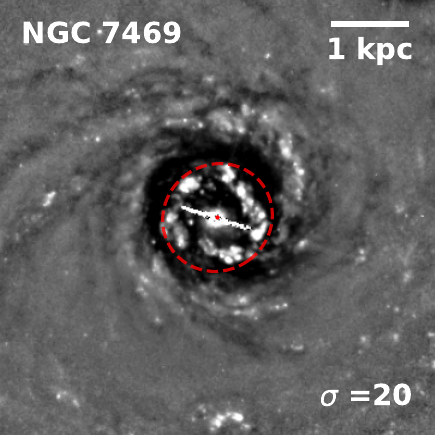}
\includegraphics[width=.24\linewidth]{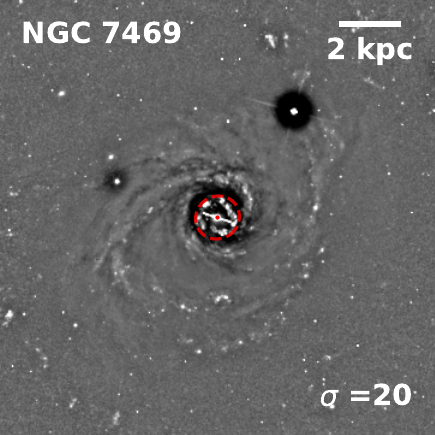}\\ 
\vspace{4.5 pt}

 \includegraphics[width=.24\linewidth]{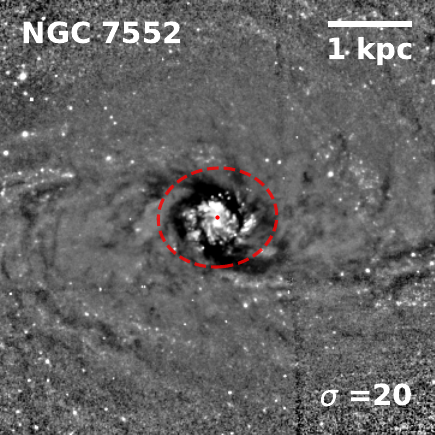}
\includegraphics[width=.24\linewidth]{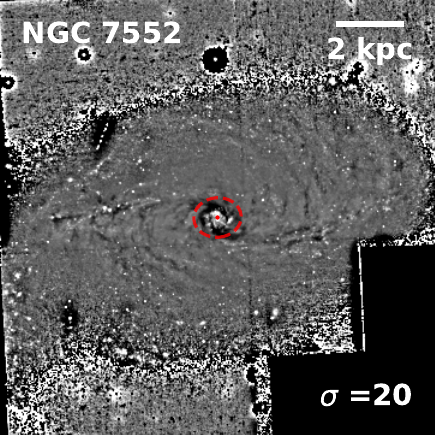}
\hspace{0.8 pt}
 \includegraphics[width=.24\linewidth]{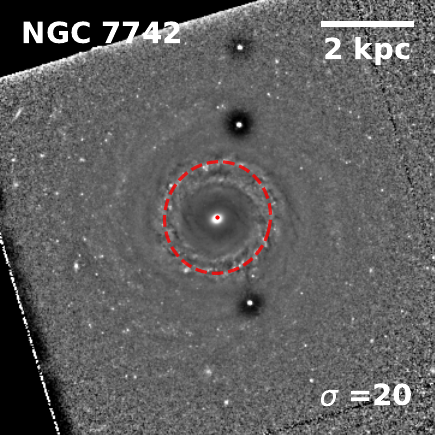}
\includegraphics[width=.24\linewidth]{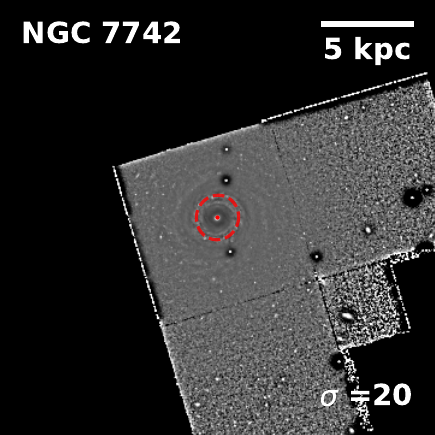}\\ 
\vspace{4.5 pt}
\caption{Continued.}
\end{figure*}

\end{appendix}
\end{document}